\documentclass{emulateapj}
\usepackage{longtable, hyperref, graphicx, subfigure, epstopdf}
\usepackage{amsmath}
\usepackage{lmodern}
\usepackage[T1]{fontenc}
\usepackage{multirow}
\usepackage{booktabs}

\usepackage{verbatim}
\usepackage[space]{grffile}
\usepackage{fancyref}
\usepackage[utf8]{inputenc}
\usepackage{url}
\usepackage{amsfonts,amsmath,amssymb}
\usepackage{latexsym}
\usepackage{natbib}

\newcommand{\Mwd}{\ifmmode {M_{\rm WD}}\else $M_{\rm WD}$\fi}
\newcommand{\tauc}{\ifmmode {\tau_{\rm cool}}\else $\tau_{\rm cool}$\fi}
\newcommand{\Msun}{\ifmmode {M_{\odot}}\else${M_{\odot}}$\fi}
\newcommand{\Rsun}{\ifmmode {R_{\odot}}\else${R_{\odot}}$\fi}
\newcommand{\lessim }{{\lower0.8ex\hbox{$\buildrel <\over\sim$}}}
\newcommand{\gessim }{{\lower0.8ex\hbox{$\buildrel >\over\sim$}}}
\newcommand{\Teff}{\ifmmode {T_{\rm eff} }\else $T_{\rm eff}$\fi}
\newcommand{\logg}{\ifmmode {{\rm log}\ g }\else log~$g$\fi} 
\newcommand{\kms}{km s$^{-1}$}
\def\deg{\ifmmode^{\circ}\else$^{\circ}$\fi}
\def\amin{\ifmmode^{\prime}\else$^{\prime}$\fi}
\def\asec{\ifmmode^{\prime\prime}\else$^{\prime\prime}$\fi}

\def\mesa{{\tt MESA}}

\slugcomment{DRAFT \today}

\shorttitle{HS 2220$+$2146}

\shortauthors{Andrews et al.}
\begin{document}

\title{Today a Duo, But Once a Trio?\\The Double White Dwarf HS 2220$+$2146 May Be A Post-Blue Straggler Binary}

\author{
Jeff J. Andrews\altaffilmark{1,2},  
Marcel Ag\"ueros\altaffilmark{2},
Warren R. Brown\altaffilmark{3},
Natalie M. Gosnell\altaffilmark{4},
A.\ Gianninas\altaffilmark{5},
Mukremin Kilic\altaffilmark{5},
Detlev Koester\altaffilmark{6}
}

\altaffiltext{1}{Foundation for Research and Technology-Hellas, 71110 Heraklion, Crete, Greece}
\altaffiltext{2}{Columbia University, Department of Astronomy, 550 West 120th Street, New York, NY 10027, USA}
\altaffiltext{3}{Smithsonian Astrophysical Observatory, Cambridge, 60 Garden St, Massachusetts, 02138, USA}
\altaffiltext{4}{Department of Astronomy, The University of Texas at Austin, 2515 Speedway, Stop C1400, Austin, TX 78712, USA}
\altaffiltext{5}{Department of Physics and Astronomy, University of Oklahoma, 440 West Brooks St., Norman, OK, 73019, USA}
\altaffiltext{6}{Institut für Theoretische Physik und Astrophysik, Universität Kiel, 24098, Kiel, Germany}

\begin{abstract}
For sufficiently wide orbital separations {\it a}, the two members of a stellar binary evolve independently. This implies that in a wide double white dwarf (DWD), the more massive WD should always be produced first, when its more massive progenitor ends its main-sequence life, and should therefore be older and cooler than its companion. The bound, wide DWD HS 2220$+$2146 ($a\approx500$ AU) does not conform to this picture: the more massive WD is the younger, hotter of the pair. We show that this discrepancy is unlikely to be due to past mass-transfer phases or to the presence of an unresolved companion. Instead, we propose that HS 2220$+$2146 formed through a new wide DWD evolutionary channel involving the merger of the inner binary in a hierarchical triple system. The resulting blue straggler and its wide companion then evolved independently, forming the WD pair seen today. Although we cannot rule out other scenarios, the most likely formation channel has the inner binary merging while both stars are still on the main sequence. This provides us with the tantalizing possibility that Kozai-Lidov oscillations may have played a role in the inner binary's merger. {\it Gaia} may uncover hundreds more wide DWDs, leading to the identification of other systems like HS 2220$+$2146. There are already indications that other WD systems may have been formed through different, but related, hierarchical triple evolutionary scenarios. Characterizing these populations may allow for thorough testing of the efficiency with which KL oscillations induce stellar mergers.
\end{abstract}


\section{Introduction} \label{sec:intro}
While roughly half of all Galactic field stars have at least one stellar companion, these companions usually have little to no impact on their evolution: the separations between the stars are generally too great for them to interact. \citet{dhital2015} recently used the Sloan Digital Sky Survey \citep[SDSS;][]{york00} to identify 10$^5$ binaries whose projected physical separations peaks at $>$10$^4$ AU. For $M\lesssim8$~\Msun, stars in such widely separated binaries never interact, independently evolving through the main sequence (MS) and giant branches and becoming white dwarfs (WDs).

Wide double WDs (DWDs), the evolutionary endpoints of these wide binaries, were first identified in catalogs of nearby stars as a subset of the common proper motion pairs \citep{sanduleak82,greenstein86,sion91}. As they are difficult to find, the number of wide DWDs has historically remained small, but recently \citet{andrews15} expanded the number of candidate and confirmed wide DWDs to 142. This sample includes two spectroscopically confirmed \emph{triple} degenerate systems, Sanduleak A/B \citep{maxted00} and G 21-15 \citep{farihi05}. These systems are composed of an unresolved pair of WDs with another, widely separated, WD companion. \citet{reipurth12} argued that such hierarchical systems may be the natural evolutionary endpoint of MS triple systems: the inner binary tightens while the outer companion's orbit expands. 

Hierarchical triples with a sufficiently large mutual inclination angle are subject to dynamical instability known as Kozai-Lidov (KL) oscillations: the outer star in a hierarchical triple perturbs the inner binary, leading to large oscillations in the inner binary's eccentricity and in the relative inclination of the two orbital planes \citep{kozai62,lidov62}. \citet{harrington68} first pointed out that these large-amplitude eccentricity oscillations would lead to a decreased periastron separation of the inner binary, possibly causing tidal interactions, mass exchanges, or even stellar mergers. These ideas have since been applied to the orbital distribution of triple stellar systems and of hot Jupiters \citep{mazeh79,kiseleva98,eggleton01,eggleton06,fabrycky07,naoz11,naoz12}.

In stellar triples, when tidal forces and magnetic braking are included \citep{perets09}, or the KL equations are expanded to the octupole order \citep{naoz14}, some of the inner binaries merge, forming blue stragglers with a wide companion. Usually observed in stellar clusters as MS stars that are bluer and more luminous than the MS turn-off \citep{sandage53,johnson55,ferraro99}, blue stragglers are formed when either accretion from, or a merger with, a companion provides fresh fuel to a star extending its lifetime \citep{mccrea64, hills76}. If the component stars are massive enough to evolve off the MS in a Hubble time, a blue straggler binary formed from the merger of the inner binary in a hierarchical triple will ultimately form a DWD.

\citet{andrews15} identified HS 2220$+$2146 (hereafter HS 2220) as an unusual DWD system. The WDs are 6$\farcs$2 apart, which at the spectroscopic distance of 76~pc corresponds to a projected physical separation of 470~AU. This suggests that these two coeval WDs evolved separately. However, spectra obtained with the Ultraviolet and Visual Echelle Spectrograph (UVES) on the Very Large Telescope (VLT) indicated that the more massive WD in this system has a younger cooling age (\tauc) than its less massive companion. In an independently evolving binary, the initially more massive star should evolve into a more massive WD before its companion. Since the more massive WD is apparently the younger in this DWD, HS 2220 cannot be explained through standard binary evolution.

We show that the properties of HS 2220 are consistent with an evolutionary history in which the inner binary in a hierarchical triple merged to form a blue straggler, which then evolved into a WD. If this scenario is correct, HS 2220 would be the first DWD known to have formed through this evolutionary channel, and confirmation that hierarchical triple systems can indeed lead to stellar mergers.

In Section \ref{sec:obs}, we use spectroscopy and gravitational redshift measurements to demonstrate that the hotter WD in HS 2220 is indeed the more massive WD. In Section~\ref{sec:accretion}, we show that these two WDs are far enough apart that they evolved independently. We describe possible formation scenarios for the DWD in Section~\ref{sec:scenario}, and include a discussion of how KL oscillations might be responsible for the formation of the more massive WD via merging of an inner binary. We put HS 2220 in the greater context of triple systems and conclude in Section~\ref{sec:disc_conc}.

\section{Observations and System Characteristics} \label{sec:obs}
\citet{baxter14} first identified the two WDs in HS 2220 as an associated pair. Using the astrometry from SDSS Data Release 7 \citep{DR7paper}, these authors showed that the two WDs have a relatively small separation, matching proper motions, and similar distance moduli, and are therefore a likely wide binary. We recovered this system in \citet{andrews15} as part of our search for wide DWDs in SDSS DR9. 

\subsection{Spectroscopy and Derived Values}
Two spectra of each WD in the system were taken as part of the Supernova Progenitor surveY \citep[SPY;][]{koester09} on 2002 Sep 25 and 26 with UVES on the VLT. Based on the \citet{koester09} fits to these $R \approx 14,000$ spectra, \citet{baxter14} argued that when observational uncertainties are taken into account, the two WDs are consistent with having been born at the same time and evolving independently.\footnote{\citet{baxter14} increase the uncertainties on the original SPY survey fits for \logg\ from 0.01 to 0.07.}

\citet{baxter14} also obtained low-resolution ($R \approx$ 1000) spectra of the WDs on 2009 Jul 20 and 25 using the B600 grating with the Multi-Object Spectrograph on Gemini-North (GMOS). We provide their spectroscopic fits in Table~\ref{tab:spectra}. Interestingly, the fits to these spectra indicate that the WD with the higher $\logg$ (hereafter HS 2220B) also has a larger $\Teff$ than its companion (hereafter HS 2220A).

\begin{figure}
\begin{center}
\includegraphics[width=0.99\columnwidth]{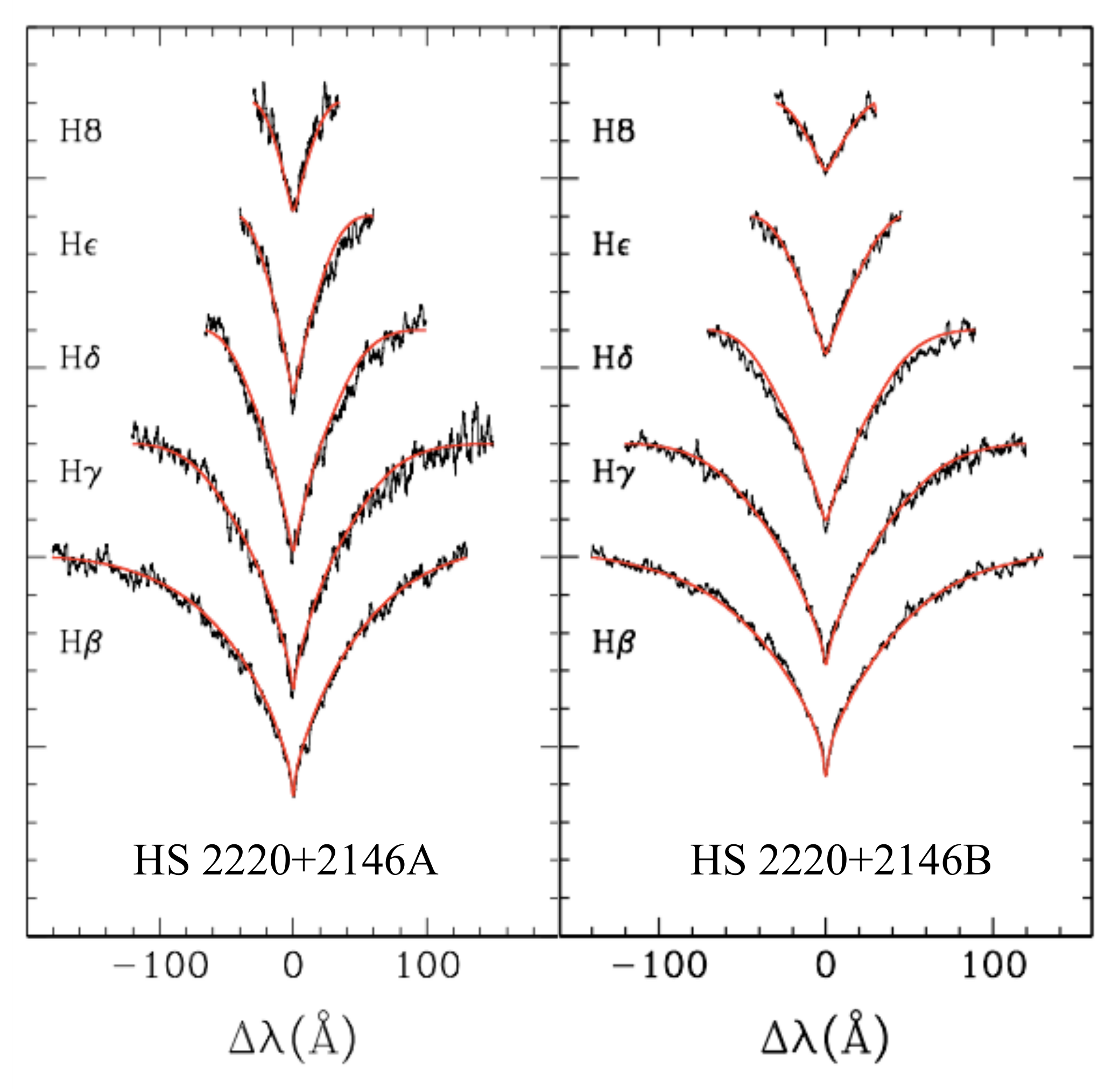}
\caption{Model fits (red lines) to the observed Balmer line profiles (black lines) of HS 2220A and B. These $R \approx 14,000$ spectra were obtained using UVES on the VLT; they are boxcar averaged with a width of 20 elements. The fitted \Teff\ and \logg\ values and the corresponding $M_{\rm WD}$ and \tauc\ are given in Table~\ref{tab:spectra}.}
\label{fig:spectra}
\end{center}
\end{figure}

We re-analyzed the UVES and archival GMOS spectra (Program ID:GN-2009B-Q-80, PI: Dobbie) in order to derive our own values for \Teff\ and \logg\ in a consistent manner.\footnote{Based on observations obtained at the Gemini Observatory acquired through the Gemini Science Archive, which is operated by the Association of Universities for Research in Astronomy, Inc., under a cooperative agreement with the NSF on behalf of the Gemini partnership: the National Science Foundation (United States), the National Research Council (Canada), CONICYT (Chile), the Australian Research Council (Australia), Ministério da Ciência, Tecnologia e Inovação (Brazil) and Ministerio de Ciencia, Tecnología e Innovación Productiva (Argentina).} Details of the UVES data reduction can be found in \citet{koester09}. We reduced the GMOS data using standard reduction techniques with the PyRAF {\tt gemini} package.\footnote{PyRAF is a product of the Space Telescope Science Institute, which is operated by AURA for NASA.}

We show the Balmer absorption lines from a UVES spectrum for each WD in Figure~\ref{fig:spectra}, along with fits to spectral model using the technique originally developed in \citet{bergeron92} and described in detail in \citet[][and references therein]{gianninas11}. These solutions are based on one-dimensional models using a mixing length parameter ML2/$\alpha = 0.8$ \citep{tremblay10, tremblay11}. The best fit \Teff\ and \logg\ values from our fits to both sets of spectra are given in Table~\ref{tab:spectra}. These spectra are an important consistency check, as they were taken with different instruments and different spectral resolutions, and yet return values for \Teff\ and \logg\ that are in close agreement. 

\begin{deluxetable*}{lc@{$\pm$}lc@{$\pm$}lr@{$\pm$}lcr@{$\pm$}lc@{$\pm$}l}
\tablecaption{Spectroscopic Fits}
\tablehead{
\colhead{} &
\multicolumn{6}{c}{Data from UVES/VLT\tablenotemark{a}} &
\colhead{} &
\multicolumn{4}{c}{Data from GMOS/Gemini-North\tablenotemark{b}} \\
\cmidrule{2-7} \cmidrule{9-12}
\colhead{} &
\multicolumn{2}{c}{\bf Our Fits} &
\multicolumn{2}{c}{\citet{koester09}} &
\multicolumn{2}{c}{Koester Updated\tablenotemark{c}} &
\colhead{} &
\multicolumn{2}{c}{ Our Fits} &
\multicolumn{2}{c}{\citet{baxter14}} \\
\colhead{} &
\multicolumn{2}{c}{} &
\multicolumn{2}{c}{fits} &
\multicolumn{5}{c}{} &
\multicolumn{2}{c}{fits}
}
\startdata
\cutinhead{HS 2220$+$2146A}
\logg & {\bf 8.151} & {\bf 0.036} & 8.080 & 0.012 & 8.213 & 0.006 &  & 8.163 & 0.052  & 8.07 & 0.07\\
\Teff & {\bf 14270} & {\bf 274} & 14601 & 32 & 14434 & 58 &   & 14274 & 434  & 13950 & 321 \\
\cutinhead{HS 2220$+$2146B}
\logg & {\bf 8.353} & {\bf 0.035} & 8.241 & 0.008 & 8.283 & 0.005 & & 8.386 & 0.044  & 8.37 & 0.07 \\
\Teff & {\bf 18833} & {\bf 218} & 18743 & 44 & 18305 & 29 & & 19469 & 284  & 19020 & 438 
\enddata
\tablecomments{A comparison of the the various spectral fits to the WDs in HS 2220 obtained with UVES/VLT and GMOS/Gemini-North spectra. All spectra show that HS 2220B has a larger \logg\ and \Teff. We use our fits to the VLT spectra throughout the remainder of this work.}
\tablenotetext{a}{Reported values are weighted averages of fits to the two spectra of each WD obtained as part of the SPY survey \citep{koester09}.}
\tablenotetext{b}{Reported values are weighted averages of fits to the three spectra of each WD obtained by \citet{baxter14}.}
\tablenotetext{c}{We refit the VLT spectra using updated versions of the models used in the original fits of \citet{koester09}.}
 \label{tab:spectra}
 \end{deluxetable*}

Our fits to the UVES spectra indicate a difference in \logg\ of 0.202$\pm$0.050 (where the uncertainty is the quadrature sum of the individual uncertainties) between HS 2220A and HS 2220B, a greater than 4-$\sigma$ difference. For a coeval DWD whose progenitors never interacted, the WD with the larger \logg\ (and therefore larger \Mwd) should have the smaller \Teff\ (and therefore larger \tauc), since the initially more massive star should have evolved into a more massive WD first. Our fits to the UVES spectra indicate that for HS 2220, the opposite is true: the {\it less} massive WD was born first. Yet, even if both WDs' \logg\ are identical, a discrepancy in the system lifetime remains since it takes a WD of this mass $>10^8$ yr to cool from 18,000 K to 14,000 K.

Why are our conclusions about these two WDs different from those of \citet{baxter14}? One answer may lie in the spectral models, which have been modified since \citet{koester09} fit the SPY spectra. In particular, \citet{tremblay09} improved Stark-broadening calculations and included some non-ideal effects in the spectral models. In addition, the numerical fitting techniques employed differ \citep[for a discussion of these differences, see][]{gianninas11,koester14}.

Using up-to-date WD spectral models, we fit the UVES spectra applying the two codes and provide both sets of spectral values in Table~\ref{tab:spectra}. This table shows that, although there are differences between the exact spectral values for each WD, all robustly show that HS 2220B has a larger \logg\ and higher \Teff\ than its companion. We interpolate between the mass-radius tables from \citet{wood95} to obtain WD mass (\Mwd) and cooling age (\tauc) measurements from the \logg\ and \Teff. Hereafter we use the derived \Mwd\ and \tauc\ measurements from the Bergeron spectral fits applied to the VLT spectra (bold quantities in Table \ref{tab:spectra}).

\begin{deluxetable}{lcc}
\tablecaption{WD Characteristics}
\tablehead{
 &
 \colhead{HS 2220$+$2146A} &
 \colhead{HS 2220$+$2146B}
}
\startdata
$\alpha$ & 22:23:01.74 & 22:23:01.64 \\
$\delta$ & $+$22:01:25.0 & $+$22:01:31.0 \\
SDSS $g$ & 16.00 & 15.59 \\
$M_{\rm WD}$ (\Msun) & 0.702$\pm$0.022 & 0.837$\pm$0.022 \\
$\tau_{\rm cool}$ (Myr) & 289$\pm$22 & 179$\pm$14 \\
$v_{\rm radial}$\tablenotemark{a} (km s$^{-1}$) & 37$\pm$6 & 56$\pm$5 \\
$v_{\rm grav}$\tablenotemark{b} (km s$^{-1}$) & 38$\pm$2 & 53$\pm$3 \\
$M_{\rm ZAMS}$\tablenotemark{c} (\Msun) & 3.3$\substack{+0.3\\-0.2}$ & 4.5$\substack{+0.4\\-0.5}$ \\
$\tau_{\rm stellar}$\tablenotemark{d} (Myr) & 360$\substack{+90\\-80}$ & 150$\substack{+60\\-30}$ \\
\cutinhead{System Characteristics}
$\theta$ (\asec) & \multicolumn{2}{c}{6.2} \\
Distance (pc) & \multicolumn{2}{c}{76$\pm$2} \\
Separation (AU) & \multicolumn{2}{c}{470}
\enddata
\tablecomments{$M_{\rm WD}$ and \tauc\ for each WD are obtained from our fits to VLT spectra; see Table~\ref{tab:spectra}.}
\tablenotetext{a}{RVs are averages measured from our FLWO spectra (see Figure~\ref{fig:rv}). The quoted uncertainties represent random errors. There is an additional systemic uncertainty of $\approx$10~km s$^{-1}$.}
\tablenotetext{b}{Gravitational redshifts are obtained from interpolations of the WD models of \citet{wood95}.}
\tablenotetext{c}{To obtain the initial stellar masses, we use the median $M_{\rm ZAMS}$ from the posterior samples of \citet{andrews15} using the listed \Mwd.}
\tablenotetext{d}{We use \mesa\ to obtain stellar lifetimes from  $M_{\rm ZAMS}$.}
\label{tab:qualities}
\end{deluxetable}

\begin{figure}
\begin{center}
\includegraphics[width=0.99\columnwidth]{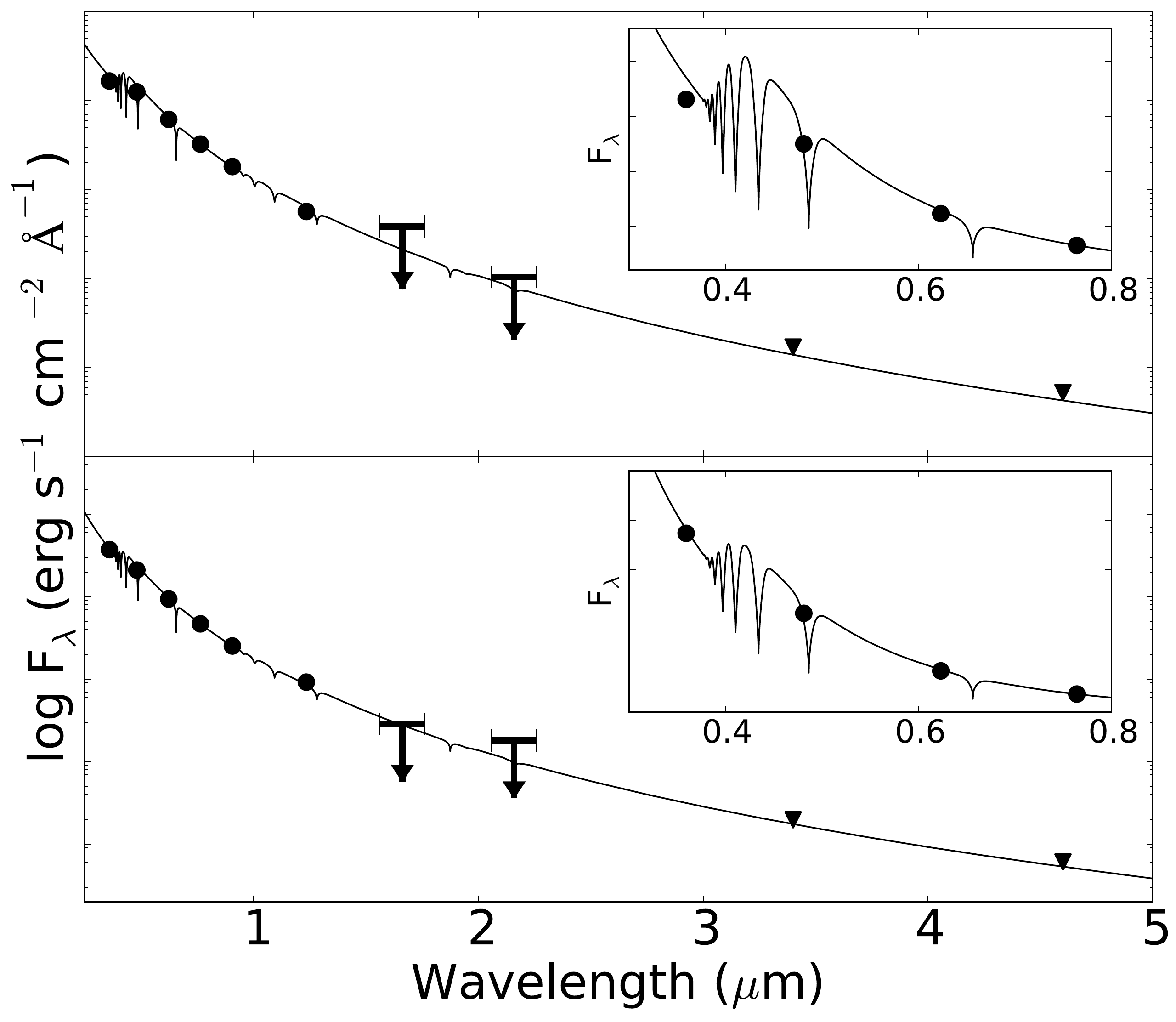}
\caption{ Using the \Teff\ values from our spectral fits, we compare the SED of HS 2220A (top panel) and B (bottom panel) with SDSS, 2MASS, and WISE photometry, normalized to the SDSS $i$ band. 2MASS $H$ and $K$ photometry only provides upper limits. WISE cannot resolve the two WDs, and the $W1$ and $W2$ photometry (triangles) is for the combined DWD system. The WISE measurements, in particular, match the predicted flux from the model WD SED, precluding the existence of a hidden late-type stellar companion. In the inset figures, we zoom in on the visible window (SDSS $ugri$) of the SED (here we plot $F_{\lambda}$, not log $F_{\lambda}$).}
\label{fig:SED}
\end{center}
\end{figure}

\subsection{Could Either WD Have a Hidden Companion?}
This conundrum could be explained by one of the WDs being in a close binary with another, unresolved star, since mass transfer could have affected both \Mwd\ and \tauc. Only a late-type dwarf or brown dwarf could escape detection in the VLT spectra. In Figure~\ref{fig:SED}, we compare the calculated spectral energy distribution (SED) derived from the best fit spectral value to SDSS, Two Micron All-Sky Survey \citep[2MASS;][]{2mass}, and Wide-field Infrared Survey Explorer \citep[WISE;][]{wise} photometry. WISE cannot resolve the DWD pair, and the $W1$ and $W2$ measurements shown in Figure~\ref{fig:SED} are therefore the combined emission from both WDs. The lack of any excess in either the 2MASS $H$ and $K$ bands (upper limits) and the WISE $W1$ and $W2$ bands eliminates the possibility of a late-type stellar companion to, or dust disk around, either WD.

\begin{figure}[]
\begin{center}
\includegraphics[width=0.99\columnwidth]{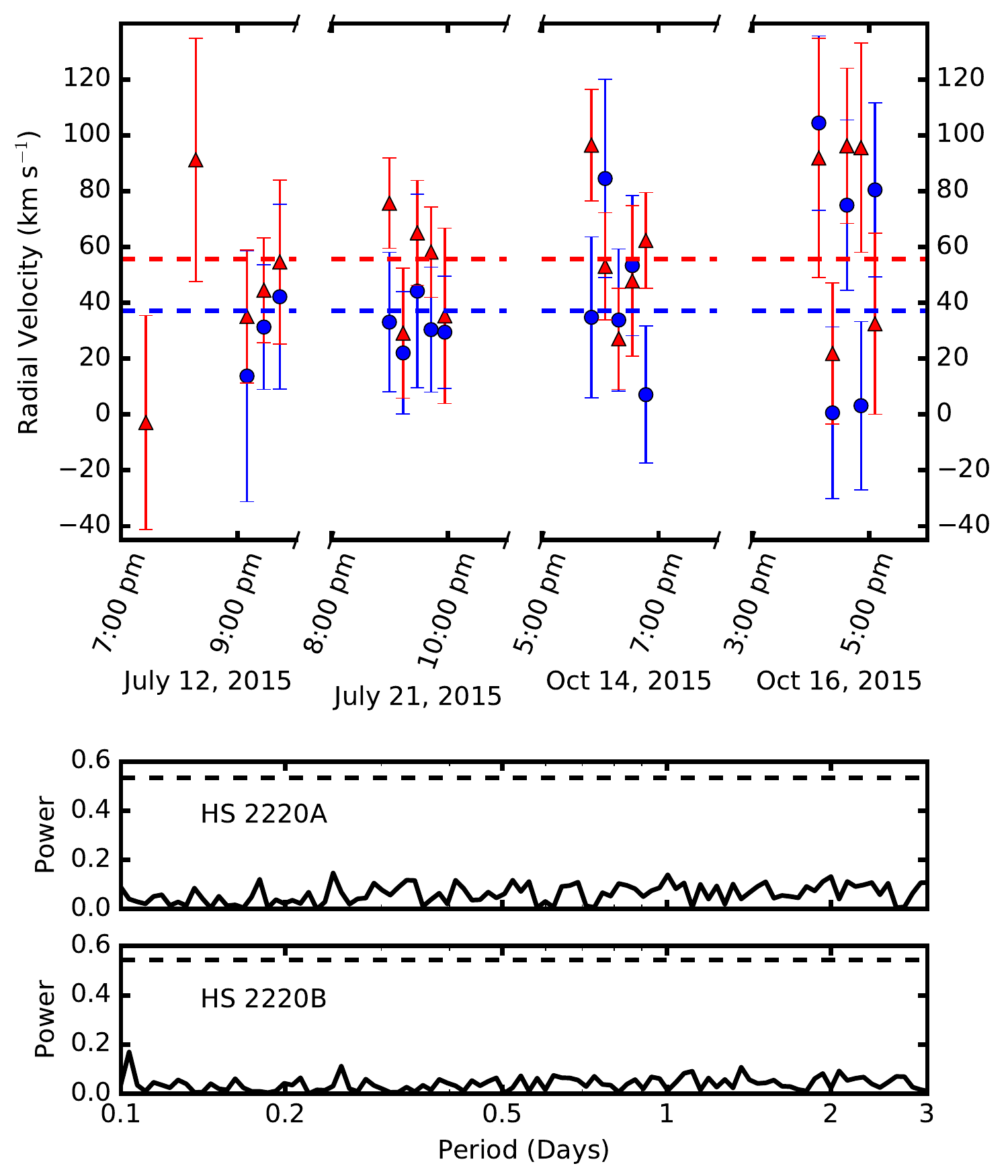}
\caption{The top panel shows RVs for HS 2220A (blue circles) and B (red triangles) calculated from the follow-up spectra taken with the FLWO 1.5-m. A positive RV corresponds to an object moving away from Earth. There is no apparent RV variation due to a hidden binary companion. HS 2220B has a larger average RV (red dashed line) than its companion (blue dashed line) because its larger mass causes a larger gravitational redshift. The apparent RV difference provides a consistency check for our spectroscopic solutions. The scatter in the first two data points taken on 2015 Jul 12 are due to a different CCD binning, which also prevented the fainter WD, HS 2220A, from being observed. The bottom two panels show Lomb-Scargle periodograms for both WDs. The dashed lines indicate the detection threshold at the 95\% confidence level; there are no significant periodic signals in the RV for either WD.}
\label{fig:rv}
\end{center}
\end{figure}

\begin{figure}[]
\begin{center}
\includegraphics[width=0.99\columnwidth]{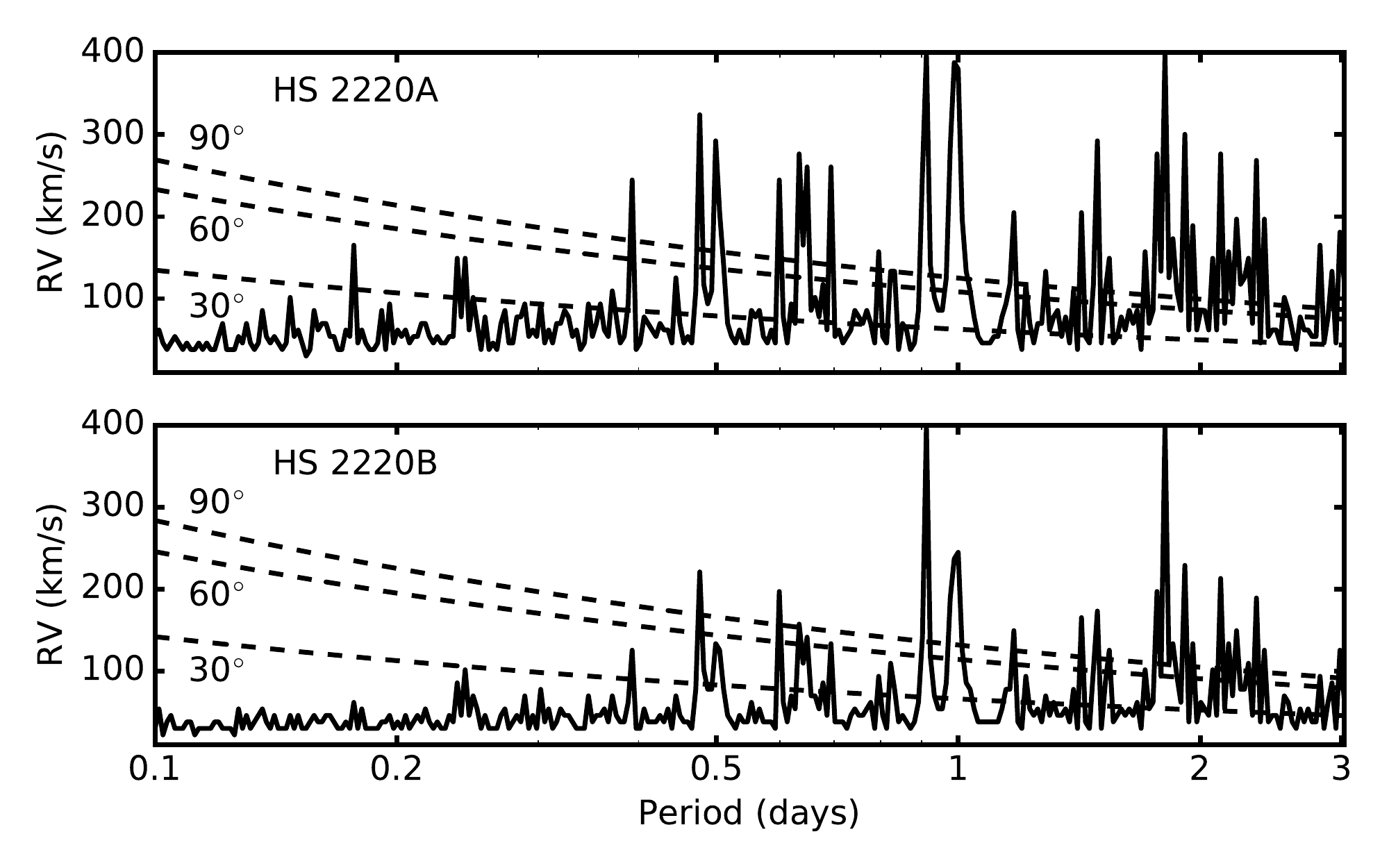}
\caption{ The solid line in each panel shows the RV constraint at the 95\% confidence level; for a particular orbital period, RVs above the line are ruled out. Assuming the limiting case that the unseen companion is a WD with a mass of 0.05 \Msun\ larger than the observed WD, we can determine the observed orbital velocities as a function of inclination angle. For nearly edge-on orbits ($i\gtrsim60^{\circ}$), orbital periods below $\sim$12 hours are ruled out for both WDs, however RVs are insensitive to an unseen companion if the inclination angle is sufficiently low.}
\label{fig:rv_limits}
\end{center}
\end{figure}

The putative hidden companion could be another WD, making the system a hierarchical triple WD. There are two known such systems, G 21-15 \citep{farihi05} and Sanduleak A/B \citep{maxted00}.\footnote{In \citet{andrews15}, we identified a third candidate triple WD system, PG 0901$+$140.} The spectrum of the unresolved DWD in both G 21-15 and Sanduleak A/B is dominated by the more luminous low-mass WD. The inverse mass-radius relation of WDs suggests that an unseen WD companion to either component of HS 2220$+$2146 would have to be more massive than the observed WD. Such an unresolved binary would have a mass substantially above the Chandrasekhar limit. Despite dedicated search programs \citep[e.g., SPY;][]{koester09}, the first super-Chandrasekhar DWD to merge within a Hubble time was only recently identified \citep{santander15}, and we therefore consider it {\it a priori} unlikely for HS 2220 to be a triple WD system.

Nevertheless, to test this possibility, we obtained additional spectra of HS2220 with the Fred Lawrence Whipple Observatory (FLWO) 1.5-m telescope in queue-scheduled time in 2015 Jul and Oct. We used the FAST long-slit spectrograph \citep{fabricant98} with the 600~line~mm$^{-1}$ grating and a 1$\farcs$5 slit, providing wavelength coverage 3500~\AA\ to 5500~\AA\ and a spectral resolution of 1.7~\AA. We rotated the slit to observe both WDs simultaneously, using the atmospheric dispersion corrector to maintain flux calibration.  Exposure times were chosen to yield a signal-to-noise (S/N) ratio of 40 per resolution element; all observations were paired with comparison lamp exposures for accurate wavelength calibration.

We maximized our sensitivity to radial velocity (RV) variability with the following approach.  We used the cross-correlation package RVSAO to measure RVs from the full spectra \citep{kurtz98}, starting with a high S/N WD template to measure absolute velocities.  We then shifted the individual spectra to rest-frame, summed them together to create templates of each WD, and cross-correlated the individual spectra with the summed templates.  The resulting precision is typically $\pm$20~\kms, limited by the large Balmer lines of these $\log{g}\approx8$ WDs.

We show the RV measurements and the mean radial velocity (horizontal dashed lines) in Figure \ref{fig:rv}. We find no evidence for significant RV variation in either WD; reduced $\chi^2$ values around the mean are 0.92 and 1.10 for HS 2220A and HS 2220B, respectively. The bottom two panels of Figure \ref{fig:rv} show Lomb-Scargle periodograms for both WDs over the expected orbital period range for close compact object binaries.\footnote{These periodograms and the 95\% confidence levels were generated using routines from the {\tt astroML} module \citep{astroML}.} The dashed line in each of these panels shows a detection at the 95\% level, calculated using a bootstrap method. There are no significant periodic signals in the RV curves for either WD.

The periodograms only determine the strength of any periodic signal. In Figure \ref{fig:rv_limits}, we show the possibility that an orbit with a particular period and RV semi-amplitude could remain unobserved by our RV observations. To determine these constraints, for each orbital period and RV semi-amplitude pair in a grid spanning the range plotted,\footnote{We choose the range based on other close compact object binary populations such as the ELM WD sample \citep[e.g.,][]{brown16} and the sample of post-common envelope binaries in SDSS \citep{nebot11}.} we use an optimization algorithm to find the best fit phase and velocity offset that minimizes the $\chi^2$ statistic when compared to the RV observations. If the resulting minimum $\chi^2$ value is above a certain threshold for $n-2$ degrees of freedom, where $n$ is the number of observations, then no orbit with the input period and RV semi-amplitude can satisfy the data. In both panels of Figure \ref{fig:rv_limits} the lines show the RV constraints at the 95\% confidence level for a given orbital period; velocities above the line are inconsistent with the RV data. 

We can convert the RV constraints to limits on binary orbits with a potential companion. In the limiting case, an unseen WD companion will have a mass just larger than the observed WD so that it produces the smallest RV variations while still avoids detection in our high resolution VLT spectra. For these putative hidden companions, we adopt masses of 0.05 \Msun\ larger than the observed WD. For the given masses and orbital period, the resulting RV semi-amplitude then depends solely on the inclination angle. The dashed lines in Figure \ref{fig:rv_limits} show how orbits with inclination angles of 30\deg, 60\deg, and 90\deg\ compare with constraints from our RV data. For orbital periods below $\sim$12 hours, nearly edge-on orbits ($i\gtrsim60^{\circ}$) are ruled out for both WDs, however longer period orbits could escape detection. Orbits with a sufficiently small inclination angle can escape detection if the resulting RV semi-amplitude is near the $\sim$20 \kms\ uncertainties on the RV measurements.

The dashed lines in Figure~\ref{fig:rv} show the average apparent RV of each WD, which we provide along with the derived uncertainty in Table~\ref{tab:qualities}. Interestingly, HS 2220B has an apparent RV 18.5$\pm$8.2~\kms\ greater than that of its companion. If the system is indeed a wide associated binary, the two WDs must have the same physical RV; the apparent difference must be due to WDs' different gravitational redshifts. 

From our spectroscopically determined $M_{\rm WD}$ measurement for each WD, we interpolate between the WD mass-radius tables from \citet{wood95} to obtain the WD radii. The gravitational redshift from the WDs' masses and radii ($R_{\rm WD}$) are:
\begin{equation}
v_{\rm grav} = \frac{G~M_{\rm WD}}{R_{\rm WD}~c}. \label{eq:v_grav}
\end{equation}
The contribution to the apparent RV from this redshift is included in Table~\ref{tab:qualities}.

These calculations provide an additional, independent consistency check on the spectroscopic results in Table~\ref{tab:spectra}: the difference in the gravitational redshifts matches the observed RV difference.

From the \Mwd\ derived from our spectral models for each star, we can determine the progenitor's zero-age main-sequence (ZAMS) mass ($M_{\rm ZAMS}$) by applying an initial-final mass relation (IFMR). Using samples from the posterior distribution of IFMRs from \citet{andrews15}, we find that HS 2220A evolved from a 3.3$\substack{+0.3\\-0.2}$~\Msun\ MS star. Uncertainties indicate the 68\% confidence level. We run a suite of stellar evolution models using \mesa\footnote{\url{ http://mesa.sourceforge.net}} \citep{paxton11,paxton13} and find that this star had a lifetime of 360$\substack{+90\\-80}$ Myr.\footnote{For details about our \mesa\ models, see \citet{andrews15}.}  Using the \citet{andrews15} IFMR again, we determine that HS 2220B came from a 4.5$\substack{+0.4\\-0.5}$~\Msun\ MS star. Our \mesa\ models indicate this star had a lifetime of 150$\substack{+60\\-30}$ Myr. Table~\ref{tab:qualities} shows our estimates for the derived MS masses and stellar lifetimes. The derived stellar lifetimes, combined with the WD cooling ages, leads to an overall age difference of $\approx$320 Myr between the two WDs.

\section{Is accretion responsible for altering the evolutionary history of HS 2220?}
\label{sec:accretion}

We wish to examine whether HS 2220 could have evolved from a primordial binary in which mass transfer either delayed the evolution of the primary or caused it to appear younger than its true age. 

\subsection{Constraining the Likely Mode of Mass Accretion in HS 2220's Putative Binary Progenitor}
We begin by using HS 2220's current projected separation of 470 AU to infer the orbital separation, $a$, during previous evolutionary states. For a widely separated binary, mass lost can be assumed to carry the specific angular momentum of the mass-losing star. This is the so-called Jeans mode mass loss:
orbital eccentricity does not change secularly, and $a (M_1 + M_2)$ is a conserved quantity \citep{hadjidemetriou63}. This is an approximation that ignores any interaction between the stellar wind and either of the stars. Interactions with the wind will tend to counteract orbital expansion due to mass loss \citep{theuns96, hurley02}; the orbital separations derived by the Jeans approximation may be underestimated.

If we take the current projected separation as a lower limit on $a$, we can estimate the orbital separation of the outer binary at birth to have been $\gtrsim$90~AU, expanding to at least 170 AU after the 4.5 \Msun\ star became a WD. The true orbital separation was likely somewhat larger due to the unknown inclination and phase of the orbit \citep{fischer92}. For these masses and separations, the Roche lobe radii ($R_L$) are dozens of AU, while the maximum asymptotic giant branch (AGB) stellar radii ($R_{\star}$) are only a few AU. $a$ was therefore large enough throughout this system's lifetime that neither star would have overfilled its Roche lobe as an AGB star.

However, either star could have accreted mass from winds generated while the other was on the AGB.  When the first star evolved into a WD, could enough mass have been accreted by its MS companion to affect that star's mass and/or evolutionary timescale? When the second star, in turn, evolved into a WD, did enough of its mass accrete onto the now-formed WD companion to affect its \logg\ and \Teff?

Wind mass loss of AGB stars is thought to occur in two stages \citep[see][and references therein]{vassiliadis93}. First, pulsations on the AGB give rise to a low-velocity wind, which is slowed by gravity as it expands and cools. When the wind temperature decreases sufficiently, dust grains condense \citep[at $\approx$1000 K for silicates and $\approx$1500 K for amorphous carbon grains;][]{hofner09}. With their increased opacity, dust grains drive a second stage: radiation pressure impacts the dust grains, which are coupled to the surrounding gas, and the wind is quickly accelerated away from the star. 

In a binary, the AGB's companion can change this picture. \citet{mohamed07} showed that a detached companion at $a \sim$ tens of AU could gravitationally focus the wind of an AGB star before it is accelerated by radiation pressure, a process termed Wind Roche Lobe Overflow \citep[WRLOF; see also][]{mohamed12}. \citet{abate13} suggested that WRLOF occurs when the dust formation radius, $R_c$, is a significant fraction of the donor star's $R_L$. Specifically, these authors found that when $R_c > 0.4~ R_L$, the accretion rate, $\dot{M}_{\rm acc}$, is enhanced with respect to the canonical Bondi-Hoyle-Littleton (BHL) $\dot{M}_{\rm acc}$, while for $R_c < 0.4~ R_L$, the accretion rate is well approximated by the BHL $\dot{M}_{\rm acc}$.

\citet{hofner09} argued that silicates and amorphous carbon grains both typically have $R_c \approx 2-3$ $R_{\star}$.\footnote{$R_c$ may be somewhat larger for silicates when there is significant iron present.} Recent observations of M-type AGB stars using mid-infrared interferometry seem to agree, showing $R_c/R_{\star} \approx 2$ for Al$_2$O$_3$ and $R_c/R_{\star} \approx 4$ for silicates \citep{karovicova13}. Additionally, \citet{olofsson02} and \citet{gonzalez_delgado03} find that they can reasonably approximate the transition line profiles of CO and SiO in several dozen AGB stars using a constant wind velocity model for distances greater than a few $R_{\star}$. 

In our \mesa\ simulations, we find that a 4 \Msun\ star reaches a maximum $R_{\star} \approx2.7$ AU, while the potential donor star has $R_L \approx 35$ AU, assuming an orbital separation of 90 AU. For all subsequent evolution, the system's separation, and hence $R_L$, is even larger. WRLOF is therefore unlikely to have operated in HS 2220, and $\dot{M}_{\rm acc}$ should be well approximated by the BHL rate.

\subsection{Estimating the Bondi-Hoyle-Littleton Accretion in HS 2220's Putative Binary Progenitor}
BHL accretion assumes a companion accreting mass from a plane-parallel wind with a constant velocity. 
We begin by determining the Bondi radius: 
\begin{equation}
r_{\rm acc} = \frac{2 G M_2}{v_{\rm wind}^2 + v_{\rm orb}^2}, \label{eq:r_bondi} 
\end{equation}
where we combine the AGB wind velocity, $v_{\rm wind}$, and the orbital velocity, $v_{\rm orb}$, to determine the relative velocity between the wind and the accretor, and $M_2$ is the mass of the accretor. A wind speed of 10 km s$^{-1}$ is typical \citep{gonzalez_delgado03}, and $M_2$ and $v_{\rm orb}$ depend on the specifics of the binary in question.

\subsubsection{Accretion while we have a AGB + MS binary}
We first consider the evolution of the 4.5 \Msun\ outer star into a 0.837 \Msun\ WD, with the 3.3 \Msun\ MS companion as the accretor at a separation of $\approx$90 AU. For this binary, Equation~\ref{eq:r_bondi} indicates $r_{\rm acc} \approx 33$~AU. 

If $\Delta M_{\rm donor} = M_{\rm ZAMS} - \Mwd \approx 3.7\ \Msun$ is the mass lost by the AGB star as it becomes a WD, we can estimate how much is accreted by the MS star in the plane parallel limit by determining the fraction of the sky subtended by the MS star's $r_{\rm acc}$ given the binary separation $a$. In the limiting case, all the mass that falls within the Bondi radius is accreted:
\begin{eqnarray}
M_{\rm acc} &\approx& \frac{\pi r_{\rm acc}^2}{4 \pi a^2} \Delta M_{\rm donor}  \label{eq:M_acc} \\
&\approx& 1.3 \times 10^{-1} \left( \frac{r_{\rm acc}}{33~ {\rm AU}} \right)^2 \left( \frac{a}{90~{\rm AU}} \right)^{-2} \left( \frac{\Delta M_{\rm donor}}{3.7~ \Msun} \right) \Msun. \nonumber
\end{eqnarray}

Here, the Bondi radius is a significant portion of the orbital separation, and the plane parallel assumption may not be appropriate, since the effect of the Roche potential may need to be taken into account. This is equivalent to the ``fast wind" assumption ($v_{\rm wind}/v_{\rm orb} >> 1$) no longer being valid. However, using three-dimensional smoothed particle hydrodynamics (SPH), \citet{theuns96} show that in a close binary when $v_{\rm wind}$ and $v_{\rm orb}$ are of the same order, the BH rate significantly overestimates the accretion rate. Later simulations by \citet{mastrodemos99} of dust-driven AGB winds in wide, detached binaries also find that BH-derived accretion rates are overestimates. We therefore consider the accretion mass given by Equation \ref{eq:M_acc} a conservative upper limit.

Even if we make the assumption that all the mass falling within $r_{\rm acc}$ is accreted, this is only $\approx$0.1 \Msun. The addition of so much mass alters the MS star's lifetime by only $\approx$20 Myr, far from enough to explain the $\approx$320 Myr age discrepancy in HS 2220.

\subsubsection{Accretion while we have a WD + AGB binary}
Next, we consider the evolution of the 3.3 \Msun\ star into a 0.702 \Msun\ WD, with the 0.837 \Msun\ WD as the accretor. Using Equation \ref{eq:r_bondi}, we determine that $r_{\rm acc} \approx 12$~AU for the WD; in this case $\Delta M_{\rm donor} \approx 2.6\ \Msun$.
With $a$ now 170 AU, we can determine the amount of mass accreted by the WD using Equation \ref{eq:M_acc}:
\begin{equation}
M_{\rm acc} \approx 3 \times 10^{-3} \left( \frac{r_{\rm acc}}{12~ {\rm AU}} \right)^2 \left( \frac{a}{170~{\rm AU}} \right)^{-2} \left( \frac{\Delta M_{\rm donor}}{2.6~ \Msun} \right) \Msun \nonumber
\end{equation}

\begin{figure}[hbt]
\begin{center}
\includegraphics[width=0.99\columnwidth]{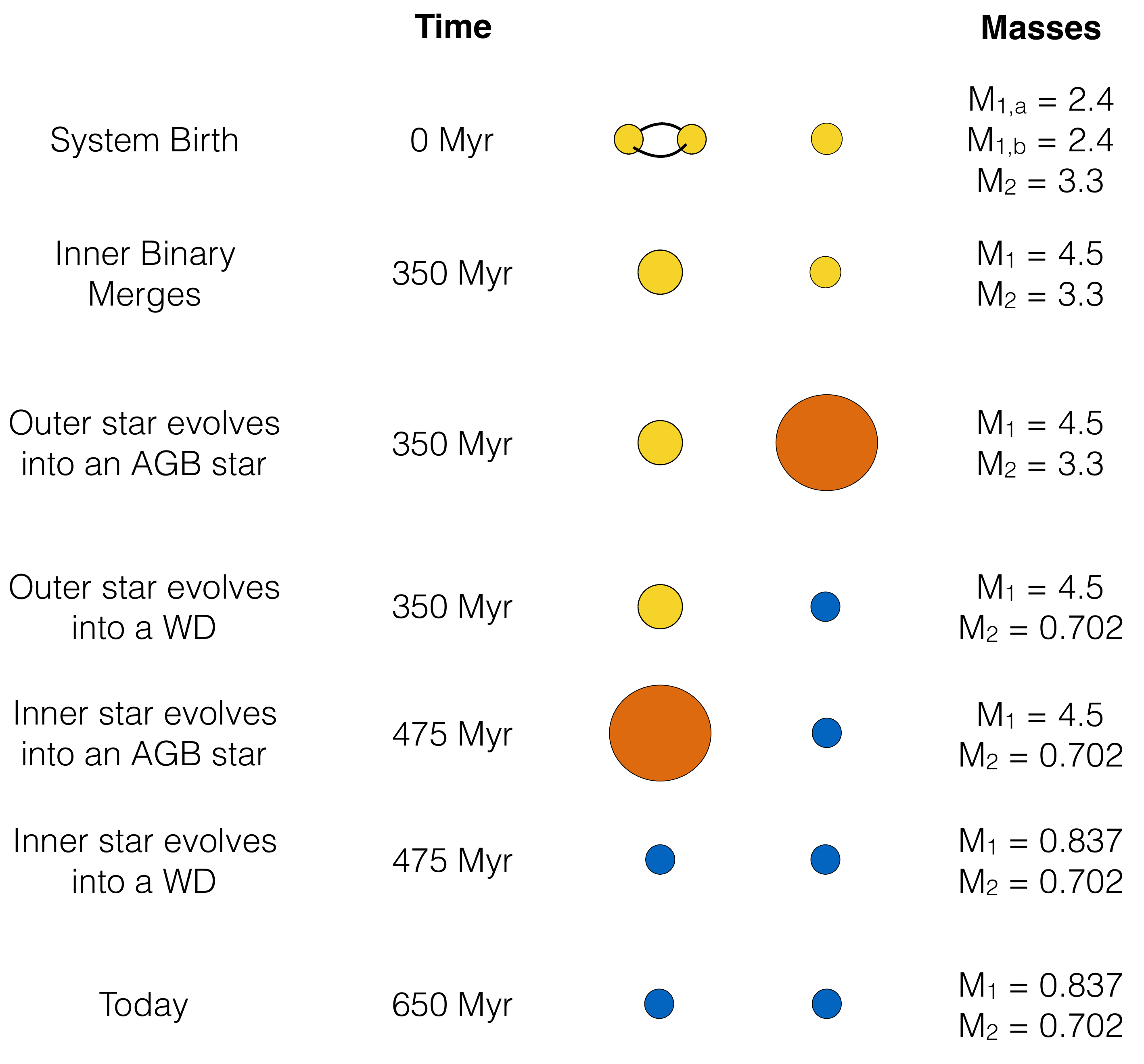}
\caption{ Our proposed formation scenario for HS 2220. The inner binary in a hierarchical triple system merged, forming a blue straggler. The two remaining stars then evolved independently, forming the WDs observed today. The timeline is inferred by working backward from the observed WD \tauc\ and stellar lifetimes assumed for the WD progenitors. These are only rough time steps, rounded to the nearest 25 Myr. }
\label{fig:scenario}
\end{center}
\end{figure}

Due to its strong dependence on $a$, which we estimated from the observed projected physical separation, this $M_{\rm acc}$ is an order of magnitude estimate at best. Still, even as little as 10$^{-5}$~\Msun\ of hydrogen accreted onto a WD is enough to induce hydrogen burning \citep[e.g.,][]{nomoto07, wolf13}.  A detailed discussion of this particular evolutionary phase is outside the scope of this work, but for the purposes of this analysis, we note that such nuclear burning would temporarily increase the WD's \Teff\ but leave its mass largely unaffected. 

Although the increased \Teff\ leads to a reduced \tauc, it cannot completely account for the cooling age discrepancy we observe in HS 2220. In the limiting case, a nova (or series of nova outbursts) efficiently heat the WD to several 10$^5$ K \citep{cunningham15}, similar to the birth temperature of a WD. However, the WD's cooling tracks remain essentially unchanged, and the two WDs would have roughly the same \tauc. To recreate the \tauc\ difference observed in HS 2220, the 0.837 \Msun\ WD would have to be steadily burning material for 10$^8$~yr, which is significantly longer than the 10$^4$ to 10$^5$~yr lifetimes of planetary nebulae \citep{badenes15}. 

Given the several conservative assumptions used to derive this accreted mass, it is possible that the system never underwent hydrogen burning. But even if hydrogen fusion did occur, our conclusion that the more massive WD is the younger one in the pair remains unaffected.

\section{Possible Formation Scenarios} \label{sec:scenario}

A star’s evolution can be delayed if it results from the merging of two MS stars. Indeed, this scenario is one evolutionary channel that may explain the blue straggler phenomenon \citep{hills76}. Simulations have shown that two merging MS stars mix, creating a higher-mass, rejuvenated merger product with a delayed evolution \citep{glebbeek08a}. Depending on the initial configuration, however, a binary may not merge until later in its evolution after one or both of the stars have evolved off the MS. In this scenario, the merger leads to a more massive WD than expected from single-star evolution. We first briefly discuss the possibility of an evolved stellar merger forming HS 2220B before returning to the scenario involving the merger of two MS stars.

\subsection{Merger of an evolved pair}

As the more massive star in a binary evolves off the MS into a giant, its radial expansion may lead to Roche lobe overflow. If unstable, the ensuing mass transfer may lead to a merger in a common envelope \citep[for a review, see][]{ivanova13}. Since the star has already left the MS, and therefore lived $\approx$90\% of its total lifetime, this merger does not significantly change the length of its life. Therefore, for a stellar merger to have led to the formation of HS 2220B in the derived time of $\approx$475 Myr, the primary must be $\approx$3 \Msun; any more massive and the star will evolve off the MS too soon, ultimately forming a WD too quickly to explain HS 2220B. 

Upon merger, the helium core of an evolved primary remains largely unaffected and its companion is absorbed into the stellar envelope \citep[cf.~discussion in][]{hurley02}. Such a merger can extend the AGB phase, since the substantially more massive H-envelope takes longer to be expelled in the stellar wind. Although insignificant compared to the overall lifetime, this extended phase can, in principle, allow the AGB core to grow. 

Estimating this growth is not straightforward: while the core can grow between pulses during the thermally pulsing AGB phase, dredge-up during the pulses can erode it, removing part or all of that added mass \citep{marigo13}. The BaSTI stellar evolution models calculate that a 3 \Msun\ star will produce a CO-core mass of 0.509\ \Msun\ and a He-core mass of 0.536 \Msun\ at the first thermal pulse \citep{pietrinferni04,pietrinferni06}.\footnote{These masses are reported for BaSTI models using scaled solar abundances, without overshoot, a metallicity of $Z=0.0198$, and Reimers parameter $\eta=0.4$. Variations on these model parameters all produce core masses within 0.1 \Msun.} This core would need to grow $\approx$0.3 \Msun\ during the thermally pulsing AGB phase to ultimately produce a system similar to HS 2220. 

Clearly, determining the amount of core growth on the AGB requires detailed modeling. However, even state-of-the-art models depend on assumptions about physics during the AGB phase, in particular prescriptions for dredge-up, convective overshooting, and the stellar wind mass-loss rate \citep{herwig00}. For instance, \citet{weiss09} use an efficient prescription for dredge-up and find that stars typically gain only a few 0.01 \Msun\ on the AGB phase. Even simulations that allow for substantial mass growth on the AGB typically show an increase of no more than $\approx$0.1 \Msun\ \citep{kalirai14}. 

A variation of this scenario may occur if the two stars have similar masses and merge while both stars are giants. In this case, the stellar cores may merge, essentially doubling the He-core mass. Using the results from \citet{pietrinferni06}, the merger of two evolved 3~\Msun\ stars would produce a star with a He-core similar to that of a 5.5 \Msun\ star. Such a star would normally produce a WD more massive than HS 2220B, although here again the AGB evolution may be different for stellar mergers. If this scenario is viable, it requires some fine tuning: the two stars must both be $\approx$3 \Msun, and the orbital separation must be large enough that the merger occurs after both stars have evolved off the MS, but not so large that the merger expels the hydrogen envelope before the cores have coalesced.

Detailed stellar evolution simulations, with particular focus on evolution during the AGB phase, are required to determine if a post-MS binary could have formed HS 2220B. While such simulations are outside the scope of this work, we consider a formation scenario involving the merger of a post-MS primary disfavored due to the amount of core growth needed for a MS secondary and the fine-tuning needed for a post-MS secondary.

\subsection{Merger of two MS stars}

In our preferred scenario, shown in Figure~\ref{fig:scenario}, HS 2220 began as a hierarchical triple system. While all three stars were still on the MS, the inner binary merged (possibly due to tidal dissipation or dynamical interaction with the outer binary which we discuss further in Section \ref{sec:KL}), forming a blue straggler. The result of such a merger would be a wide binary composed of the merger product and a MS star. This outer, now less massive, star evolved first into the 0.702 \Msun\ WD, HS 2220A, and later the blue straggler evolved into the 0.837 \Msun\ WD, HS 2220B.

Our spectroscopic analysis provided us with \tauc, the time since each star became a WD, and \Mwd, which we combined with an IFMR and a stellar lifetime function to determine the stars' pre-WD lifetimes. 
Our \mesa\ simulations indicate that the $M_{\rm ZAMS} = 3.3\ \Msun$ progenitor of HS 2220A had a pre-WD lifetime of 360 Myr. Since it evolved independently, this star sets the age of HS 2220 at $\approx$650 Myr. 

By calculating the lifetime of the 4.5 \Msun\ blue straggler, we can determine how long after formation the inner binary merged. Our \mesa\ simulations show that a 4.5 \Msun\ star has a pre-WD lifetime of $\approx$150 Myr (135 Myr on the MS and 15 Myr in post-MS evolution), which would indicate that the inner binary in the hierarchical triple merged $\approx$325 Myr after the system's formation. However, since its progenitors had already been burning hydrogen for hundreds of Myr, the merger product had a somewhat shorter lifetime than a $M_{\rm ZAMS} = 4.5\ \Msun$ star. Calculating the appropriate lifetime for a merger product depends on the individual masses of the progenitors and what fraction of their lifetimes they have been burning hydrogen. During the merger, up to a few percent of the total binary mass is lost \citep{lombardi02}. The 4.5 \Msun\ progenitor of HS 2220B would have begun as a binary with a total mass closer to 4.8 \Msun. For demonstrative purposes, we will assume the binary was composed of two stars of 2.4 \Msun\ each.

We use our pre-merger time estimate of 325 Myr as a starting point. For two merging 2.4 \Msun\ stars that have been already burning hydrogen for 325 Myr (roughly half their expected MS lifetimes), the merger product is born already $\approx$20\% through its MS evolution \citep{glebbeek08b}. This means that the progenitor of HS 2220B existed for $\approx$125 Myr after the merger and before it became a WD. When we include this in our formation scenario, we find that the inner binary survived roughly 350 Myr before merging. We suggest that it is coincidental that this time matches the pre-WD lifetime for HS 2220A, since these times are only approximate and the two stars should be independently evolving.

Although the scenario outlined in Figure~\ref{fig:scenario} is robust to different choices of stellar lifetime function or IFMR, the timeline is suggestive, and we therefore rounded the time steps to the nearest 25 Myr.

\subsection{The role of Kozai-Lidov oscillations} \label{sec:KL}

When the orbits have a sufficiently large mutual inclination, KL oscillations cause a secular exchange of angular momentum between the inner and outer binary in hierarchical triple systems \citep{kozai62, lidov62}. \citet{perets09} showed that KL oscillations, when combined with tidal dissipation and magnetic braking, are one evolutionary channel for forming blue stragglers. In particular, these authors suggested that massive, luminous blue stragglers in the field may have massive companions that have evolved into WDs in wide orbits with long KL timescales. 

\citet{naoz14} found that expanding the KL equations to the octopolar term (the eccentric KL mechanism) led to instabilities in KL oscillations that could cause the inner binary to merge; magnetic braking is not needed \citep[for a review, see][]{naoz16}. Specifically, \citet{naoz14} find that the eccentric KL mechanism forms blue stragglers in binaries with a $P_{\rm orb}$ distribution that peaks at $\approx$ $10^5$ days (see their Figure 8). This corresponds to a separation of $\approx$100 AU, in agreement with the post-merger separation we estimate in Section~\ref{fig:scenario}. 

Since wide binaries at these separations are stable for Gyr against disruption by the Galactic tide and external perturbers \citep{jiang10}, it seems inevitable that sufficiently massive blue stragglers in such systems would evolve into wide DWDs. Using the timeline and evolutionary scenario shown in Figure~\ref{fig:scenario}, we examine whether HS 2220 could have formed in such a manner.

If we assume that HS 2220 was formed through KL oscillations, we can set a lower limit on the separation of the inner binary: the quadrupolar timescale for KL oscillations ($t_{\rm quad}$) must be shorter than the inner binary's timescale for GR precession ($t_{\rm 1,GR}$). These timescales are provided by \citet{naoz13b}:
\begin{eqnarray}
t_{\rm quad} &\approx& 2 \pi \frac{a_2^3 (1-e_2^2)^{3/2} \sqrt{M_{1,a} + M_{1,b}}}{a_1^{3/2}M_2 \sqrt{G}}, \label{eq:t_quad} \\
t_{\rm 1,GR} &\approx& 2 \pi \frac{a_1^{5/2} c^2 (1-e_1^2)}{3 G^{3/2} (M_{1,a} + M_{1,b})^{3/2}} \label{eq:t_GR},
\end{eqnarray}
where $c$ is the speed of light, $a_1$ and $e_1$ are the orbital separation and eccentricity of the inner binary, respectively, and $a_2$ and $e_2$ are the orbital separation and eccentricity for the outer binary. Since $t_{\rm quad} < t_{\rm 1,GR}$, we obtain:
\begin{equation}
a_1^4 > \frac{3 a_2^3 G}{c^2} \frac{(1 - e_2^2)^{3/2}}{1 - e_1^2} \frac{(M_{1,a} + M_{1,b})^2}{M_2} \label{eq:A_1_lim_GR}.
\end{equation}

We can set an upper limit on $a_1$ by noting that a widely separated inner binary is unstable on a short timescale. We use the constraint from \citet{naoz14} that a stable system must have $\epsilon\ <\ 0.1$, where:
\begin{equation}
\epsilon = \frac{a_1}{a_2} \frac{e_2}{1 - e_2^2}. \label{eq:A_1_lim_epsilon}
\end{equation}

\begin{figure}[]
\begin{center}
\includegraphics[width=0.99\columnwidth]{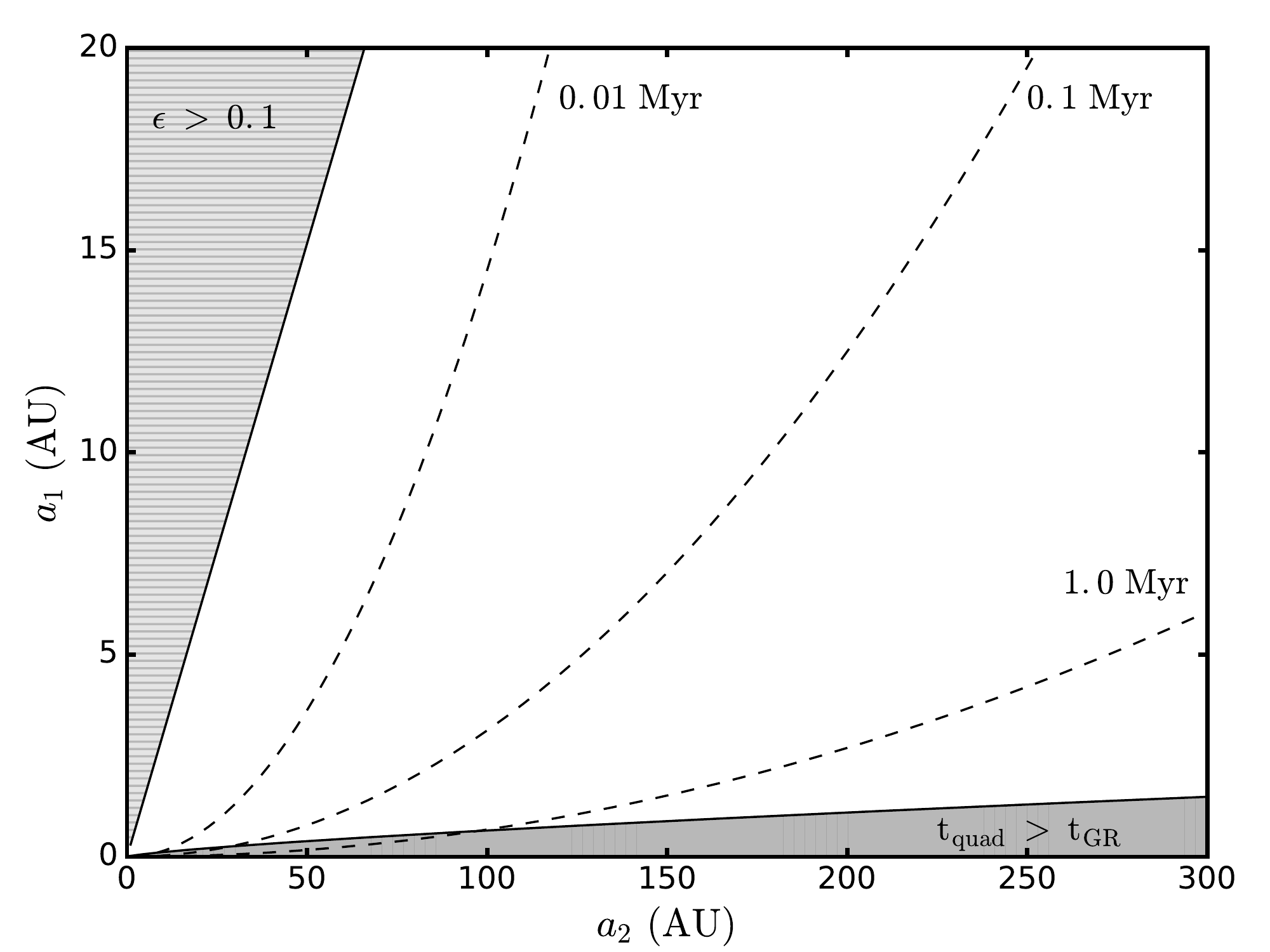}
\caption{$a_1$ as a function of $a_2$ for HS 2220's progenitor system. The lower limit on $a_1$ comes from the requirement that KL oscillations must occur on a shorter timescale than the inner orbit's GR precession. The upper limit reflects the requirement that HS 2220's progenitor begin in a stable configuration. The exact region set by these constraints depends somewhat on $e_1$ and $e_2$; we adopted $e_1 = 0.0$ and $e_2 = 0.3$. For an outer binary separation of 100 AU, the inner binary must have started with a separation of a few to tens of AU. Dashed lines show lines of constant $t_{\rm quad}$. }
\label{fig:KL_limits}
\end{center}
\end{figure}

Figure \ref{fig:KL_limits} shows the resulting allowed region of parameter space for $a_1$ and $a_2$. $a_1$ must have had a separation of a few to tens of AU (we have assumed that $e_1 = 0$ and $e_2 = 0.3$). GR precession impedes KL oscillations for smaller values of $a_1$, and the triple is not stable on short times for larger $a_1$. We also show lines of constant quadrupolar timescale; for an initial binary with $a_1\sim$~AU and $a_2 = 100$ AU, the binary had a quadrupolar timescale of $\sim$0.1 Myr.

In their parameter space study, \citet{naoz14} found that the inner binaries of hierarchical triple stars tended to merge in $\approx$4\% of systems, and within 5$-$100 $t_{\rm quad}$. Combining the adopted $t_{\rm quad}$ of 0.1 Myr with a merger time of 350 Myr, we find that the progenitor of HS 2220 would have merged after a few $10^2-$10$^3$ quadrupolar timescales, significantly longer than what was found by \citet{naoz14}. However, if the outer binary had an initial separation closer to 250 AU, the system could have had a $t_{\rm quad}$ of a few Myr, in agreement with the \citet{naoz14} prediction.

The evolution of the progenitor of HS 2220 could be further complicated if $a_1$ were a few AU at birth: \citet{naoz13b} demonstrated that in systems with $t_{\rm quad} \approx t_{\rm 1,GR}$, a resonance-like behavior between KL oscillations and GR precession is possible. A complete understanding of such an interaction and its applicability to the progenitor of HS 2220 would require detailed dynamical evolution simulations which are outside the scope of this work. Nevertheless, Figure \ref{fig:KL_limits} shows that there is a large range of parameter space in which a triple system could have been affected by KL oscillations, ultimately leading to a wide DWD binary similar to HS 2220.

\subsection{Caveats}
Our analysis depends on determining $M_{\rm ZAMS}$ for the progenitors of the WDs in HS 2220. We used the IFMR from \citet{andrews15}, which is calibrated to wide DWDs with $2\ \Msun \leq M_{\rm ZAMS} \leq 4\ \Msun$. If instead, we use the \citet{williams09} IFMR, the two WDs have somewhat lower $M_{\rm ZAMS}$, and hence longer pre-WD lifetimes, resulting in an overall system age of 875 Myr. The qualitative evolution of the system, however, is unaffected.  

The pre-WD lifetime of the blue straggler depends on a number of unknowns. We do not know how much mass was lost in the merger and therefore do not know the total mass of the inner binary at formation. Nor do we know if the inner binary contained roughly equal mass components, as we assumed, or had a smaller mass ratio, which would decrease the blue straggler's lifetime \citep{glebbeek08b}. Furthermore, given its altered composition and increased rotation, it is not clear that a blue straggler follows the same IFMR as single stars.

Finally, both the amount of mass lost in the merger and the blue straggler's lifetime are dependent on the details of the merger. \citet{glebbeek08b} assumed a head-on collision; however, in their SPH simulations, \citet{sills05} showed that the high angular momentum of a blue straggler formed from a somewhat oblique stellar merger will induce mixing, extending the lifetime relative to the head-on case. The timeline in Figure~\ref{fig:scenario} should therefore be considered  demonstrative only, and we have not included uncertainties on the time steps.

\section{Discussion and Conclusions} \label{sec:disc_conc}

HS 2220$+$2146 is a wide DWD in which the more massive WD is the younger WD in the pair, contradicting the standard expectation of stellar evolution theory.\footnote{The more massive WD in the DWD SDSS J1257$+$5428 \citep{badenes09,marsh11} was shown to have a smaller \tauc\ than its companion \citep{bours15}. However, its relatively short $P_{\rm orb} = 4.6$ hr indicates that the two WDs almost certainly interacted in the past, and it is not clear that this system followed, even qualitatively, a similar evolutionary pathway to HS 2220's.} Multiband photometry and a lack of RV variations rule out the possibility of an unseen companion in the system, and the binary is too widely separated to have had any prior mass accretion that could account for the observed $M_{\rm WD}$ and $\tau_{\rm cool}$ discrepancy.

To explain the peculiar characteristics of this system, we suggest that HS 2220 went through an alternative evolutionary channel for a DWD in which it formed as a triple system. The inner binary then merged to create a blue straggler in a wide binary, a process that could have been mediated by KL oscillations. When we reconstruct the evolutionary history of HS 2220, we find that there is a large region of parameter space in which KL oscillations are active. An inner binary initially separated by $\sim$10 AU may have merged after a few 10$^3-10^4~t_{\rm quad}$, a somewhat longer timescale for the merger than that obtained in simulations by \citet{naoz14}. Alternatively, for inner separations closer to 1~AU, the inner binary's merger may have involved a resonance-like interaction between KL oscillations and GR precession.

KL oscillations, combined with dissipation, have already been demonstrated to lead to the observed excess of hierarchical triple systems with an inner binary $P_{\rm orb}\ \lessim\ 7$ days \citep{tokovinin02,tokovinin06,fabrycky07}. A fraction of the inner binaries in triples can merge, forming blue stragglers in wide binaries, when either dissipation due to magnetic braking \citep{perets09} or higher order terms in the KL equations \citep{naoz14} are included. Sufficiently massive wide binaries will then evolve into DWDs, and since the blue straggler has a longer stellar lifetime than its mass would suggest, the two WDs will have an age discrepancy similar to the one observed in HS 2220. Although on-going studies of binary evolution may reveal other, exotic evolutionary channels that can form HS 2220, our proposed scenario naturally explains the observations of the system. 

HS 2220 is not the first WD binary suggested to have formed from a triple system. The WD in the eclipsing WD+dK2 binary V471 Tauri is the most massive known in the Hyades open cluster, but is also the hottest and youngest. \citet{obrien01} suggested that it is the end product of a blue straggler star that formed from the merger of the inner binary in a hierarchical triple. In their paradigm, however, the outer star was close enough to the inner binary that when the blue straggler evolved into a giant, the system underwent a common envelope, and the orbit shrank to the present 12.5 hr period. V471 Tauri could be the end result of one triple system evolutionary channel; compared to the progenitor of HS 2220, V471 Tauri formed from a lower-mass, outer companion in a tighter orbit.

In the opposite regime, triple systems in which the inner binary is wider may never merge but still completely evolve into WDs. The Sanduleak A/B system \citep{maxted00} and G 21-15 \citep{farihi05} are two examples of spectroscopically confirmed triple WD systems that may have formed through this evolutionary pathway. These systems are composed of spectroscopic DWD binaries with common proper motion companions. Considering their hierarchical nature, and depending on the unknown mutual inclination between the two orbital planes, it is possible that these systems, too, were effected by KL oscillations. 

The WD systems V471 Tauri, Sanduleak A/B, G 21-15, and HS 2220$+$2146 are evidence for a still largely unexplored variety of higher-order evolutionary channels forming stellar remnants. These systems may represent the vanguard of large populations of triple systems: several authors have demonstrated that spectroscopic binaries have a high likelihood of containing a tertiary companion \citep[e.g.,][]{tokovinin06}, and \citet{raghavan10} reported that some 10\% of nearby solar-type stars are in triple or higher-order systems.

Binary evolution studies, particularly with respect to compact objects, have traditionally ignored the complexities introduced by a third companion. However, considering the ubiquity of triple systems in the nearby Galactic neighborhood, it may not be safe to ignore the effect of a wide companion on a close binary, particularly with respect to rare astrophysical phenomena \citep[e.g., for a discussion of how WD triples may be an important type Ia supernova progenitor, see][]{kushnir13}.

Regardless of the application, gaining a better understanding of KL oscillations is important. HS 2220 provides a unique setting for advancing toward this goal because we can estimate the merger time of the inner binary. This system, and others like it that are not yet identified, may provide new observations with which to compare simulations. 

We were confident that HS 2220 was a curious system because of the quality of our spectra. Although this DWD is relatively nearby, these spectra did require large-diameter telescopes. In \citet{andrews15}, we identified five other candidate DWDs in which the more massive WD may be the younger one. In all those cases, higher S/N spectra are needed to confirm the discrepancies in the fitted $\tauc$ values. Even if these systems prove to have formed through a standard binary channel, it is likely that more WD systems with complicated evolutionary histories exist in our Galactic neighborhood. 

However, had HS 2220 formed 1 Gyr earlier, we likely would not have been able to identify the two WDs as having different ages: the $\tauc$ uncertainties would have been too large. In searching for more of these discrepant systems, one should therefore focus on nearby, hot wide DWDs. The upcoming {\it Gaia} data are expected to contain $\sim 10^5-10^6$ new WDs with proper motions \citep{carrasco14}. Combining the temperatures and astrometric distances obtained from {\it Gaia}'s observations with a WD mass-radius relation will result in a mass and age estimate for every one of these WDs. Identifying populations of wide DWDs in which the more massive WD is the younger may therefore be possible without spectroscopic follow up of every DWD.

\acknowledgments

We thank the anonymous referee for helpful suggestions which greatly improved this manuscript. We thank Phil Macias, Smadar Naoz, Enrico Ramirez-Ruiz, and Tassos Fragos for useful conversations about this system. M.A.A., M.K., and A.G.~gratefully acknowledge the support of the NSF and NASA under grants AST-1255419, AST-1312678, and NNX14AF65G, respectively. N.M.G.~acknowledges the support of the W.~J.~McDonald Postdoctoral Fellowship.

Funding for the Sloan Digital Sky Survey IV has been provided by
the Alfred P. Sloan Foundation, the U.S. Department of Energy Office of
Science, and the Participating Institutions. SDSS-IV acknowledges
support and resources from the Center for High-Performance Computing at
the University of Utah. The SDSS web site is www.sdss.org.

SDSS-IV is managed by the Astrophysical Research Consortium for the 
Participating Institutions of the SDSS Collaboration including the 
Brazilian Participation Group, the Carnegie Institution for Science, 
Carnegie Mellon University, the Chilean Participation Group, the French Participation Group, Harvard-Smithsonian Center for Astrophysics, 
Instituto de Astrof\'isica de Canarias, The Johns Hopkins University, 
Kavli Institute for the Physics and Mathematics of the Universe (IPMU) / 
University of Tokyo, Lawrence Berkeley National Laboratory, 
Leibniz Institut f\"ur Astrophysik Potsdam (AIP),  
Max-Planck-Institut f\"ur Astronomie (MPIA Heidelberg), 
Max-Planck-Institut f\"ur Astrophysik (MPA Garching), 
Max-Planck-Institut f\"ur Extraterrestrische Physik (MPE), 
National Astronomical Observatory of China, New Mexico State University, 
New York University, University of Notre Dame, 
Observat\'ario Nacional / MCTI, The Ohio State University, 
Pennsylvania State University, Shanghai Astronomical Observatory, 
United Kingdom Participation Group,
Universidad Nacional Aut\'onoma de M\'exico, University of Arizona, 
University of Colorado Boulder, University of Oxford, University of Portsmouth, 
University of Utah, University of Virginia, University of Washington, University of Wisconsin, 
Vanderbilt University, and Yale University.

This publication makes use of data products from the Two Micron All-Sky Survey, which was a joint project of the University of Massachusetts and the Infrared Processing and Analysis Center/California Institute of Technology, funded by the National Aeronautics and Space Administration and the National Science Foundation.

This publication makes use of data products from the Wide-field Infrared Survey Explorer, which is a joint project of the University of California, Los Angeles, and the Jet Propulsion Laboratory/California Institute of Technology, funded by the National Aeronautics and Space Administration.

\setlength{\baselineskip}{0.6\baselineskip}
\bibliography{references}

\begin{thebibliography}{87}
\expandafter\ifx\csname natexlab\endcsname\relax\def\natexlab#1{#1}\fi

\bibitem[{{Abate} {et~al.}(2013){Abate}, {Pols}, {Izzard}, {Mohamed}, \& {de
  Mink}}]{abate13}
{Abate}, C., {Pols}, O.~R., {Izzard}, R.~G., {Mohamed}, S.~S., \& {de Mink},
  S.~E. 2013, \aap, 552, A26

\bibitem[{{Abazajian} {et~al.}(2009){Abazajian}, {Adelman-McCarthy},
  {Ag{\"u}eros}, {Allam}, {Allende Prieto}, {An}, {Anderson}, {Anderson},
  {Annis}, {Bahcall}, \& et~al.}]{DR7paper}
{Abazajian}, K.~N. {et~al.} 2009, \apjs, 182, 543

\bibitem[{{Andrews} {et~al.}(2015){Andrews}, {Ag{\"u}eros}, {Gianninas},
  {Kilic}, {Dhital}, \& {Anderson}}]{andrews15}
{Andrews}, J.~J., {Ag{\"u}eros}, M.~A., {Gianninas}, A., {Kilic}, M., {Dhital},
  S., \& {Anderson}, S.~F. 2015, ArXiv e-prints

\bibitem[{{Badenes} {et~al.}(2015){Badenes}, {Maoz}, \&
  {Ciardullo}}]{badenes15}
{Badenes}, C., {Maoz}, D., \& {Ciardullo}, R. 2015, \apjl, 804, L25

\bibitem[{{Badenes} {et~al.}(2009){Badenes}, {Mullally}, {Thompson}, \&
  {Lupton}}]{badenes09}
{Badenes}, C., {Mullally}, F., {Thompson}, S.~E., \& {Lupton}, R.~H. 2009,
  \apj, 707, 971

\bibitem[{{Baxter} {et~al.}(2014){Baxter}, {Dobbie}, {Parker}, {Casewell},
  {Lodieu}, {Burleigh}, {Lawrie}, {K{\"u}lebi}, {Koester}, \&
  {Holland}}]{baxter14}
{Baxter}, R.~B. {et~al.} 2014, \mnras, 440, 3184

\bibitem[{{Bergeron} {et~al.}(1992){Bergeron}, {Saffer}, \&
  {Liebert}}]{bergeron92}
{Bergeron}, P., {Saffer}, R.~A., \& {Liebert}, J. 1992, \apj, 394, 228

\bibitem[{{Bours} {et~al.}(2015){Bours}, {Marsh}, {G{\"a}nsicke}, {Tauris},
  {Istrate}, {Badenes}, {Dhillon}, {Gal-Yam}, {Hermes}, {Kengkriangkrai},
  {Kilic}, {Koester}, {Mullally}, {Prasert}, {Steeghs}, {Thompson}, \&
  {Thorstensen}}]{bours15}
{Bours}, M.~C.~P. {et~al.} 2015, \mnras, 450, 3966

\bibitem[{{Brown} {et~al.}(2016){Brown}, {Gianninas}, {Kilic}, {Kenyon}, \&
  {Allende Prieto}}]{brown16}
{Brown}, W.~R., {Gianninas}, A., {Kilic}, M., {Kenyon}, S.~J., \& {Allende
  Prieto}, C. 2016, \apj, 818, 155

\bibitem[{{Carrasco} {et~al.}(2014){Carrasco}, {Catal{\'a}n}, {Jordi},
  {Tremblay}, {Napiwotzki}, {Luri}, {Robin}, \& {Kowalski}}]{carrasco14}
{Carrasco}, J.~M., {Catal{\'a}n}, S., {Jordi}, C., {Tremblay}, P.-E.,
  {Napiwotzki}, R., {Luri}, X., {Robin}, A.~C., \& {Kowalski}, P.~M. 2014,
  \aap, 565, A11

\bibitem[{{Cunningham} {et~al.}(2015){Cunningham}, {Wolf}, \&
  {Bildsten}}]{cunningham15}
{Cunningham}, T., {Wolf}, W.~M., \& {Bildsten}, L. 2015, \apj, 803, 76

\bibitem[{{Dhital} {et~al.}(2015){Dhital}, {West}, {Stassun}, {Schluns}, \&
  {Massey}}]{dhital2015}
{Dhital}, S., {West}, A.~A., {Stassun}, K.~G., {Schluns}, K.~J., \& {Massey},
  A.~P. 2015, \aj, 150, 57

\bibitem[{{Eggleton} \& {Kiseleva-Eggleton}(2001)}]{eggleton01}
{Eggleton}, P.~P., \& {Kiseleva-Eggleton}, L. 2001, \apj, 562, 1012

\bibitem[{{Eggleton} \& {Kisseleva-Eggleton}(2006)}]{eggleton06}
{Eggleton}, P.~P., \& {Kisseleva-Eggleton}, L. 2006, \apss, 304, 75

\bibitem[{{Fabricant} {et~al.}(1998){Fabricant}, {Cheimets}, {Caldwell}, \&
  {Geary}}]{fabricant98}
{Fabricant}, D., {Cheimets}, P., {Caldwell}, N., \& {Geary}, J. 1998, \pasp,
  110, 79

\bibitem[{{Fabrycky} \& {Tremaine}(2007)}]{fabrycky07}
{Fabrycky}, D., \& {Tremaine}, S. 2007, \apj, 669, 1298

\bibitem[{{Farihi} {et~al.}(2005){Farihi}, {Becklin}, \&
  {Zuckerman}}]{farihi05}
{Farihi}, J., {Becklin}, E.~E., \& {Zuckerman}, B. 2005, \apjs, 161, 394

\bibitem[{{Ferraro} {et~al.}(1999){Ferraro}, {Paltrinieri}, {Rood}, \&
  {Dorman}}]{ferraro99}
{Ferraro}, F.~R., {Paltrinieri}, B., {Rood}, R.~T., \& {Dorman}, B. 1999, \apj,
  522, 983

\bibitem[{{Fischer} \& {Marcy}(1992)}]{fischer92}
{Fischer}, D.~A., \& {Marcy}, G.~W. 1992, \apj, 396, 178

\bibitem[{{Gianninas} {et~al.}(2011){Gianninas}, {Bergeron}, \&
  {Ruiz}}]{gianninas11}
{Gianninas}, A., {Bergeron}, P., \& {Ruiz}, M.~T. 2011, \apj, 743, 138

\bibitem[{{Glebbeek} \& {Pols}(2008)}]{glebbeek08b}
{Glebbeek}, E., \& {Pols}, O.~R. 2008, \aap, 488, 1017

\bibitem[{{Glebbeek} {et~al.}(2008){Glebbeek}, {Pols}, \&
  {Hurley}}]{glebbeek08a}
{Glebbeek}, E., {Pols}, O.~R., \& {Hurley}, J.~R. 2008, \aap, 488, 1007

\bibitem[{{Gonz{\'a}lez Delgado} {et~al.}(2003){Gonz{\'a}lez Delgado},
  {Olofsson}, {Kerschbaum}, {Sch{\"o}ier}, {Lindqvist}, \&
  {Groenewegen}}]{gonzalez_delgado03}
{Gonz{\'a}lez Delgado}, D., {Olofsson}, H., {Kerschbaum}, F., {Sch{\"o}ier},
  F.~L., {Lindqvist}, M., \& {Groenewegen}, M.~A.~T. 2003, \aap, 411, 123

\bibitem[{{Greenstein}(1986)}]{greenstein86}
{Greenstein}, J.~L. 1986, \aj, 92, 867

\bibitem[{{Hadjidemetriou}(1963)}]{hadjidemetriou63}
{Hadjidemetriou}, J.~D. 1963, Icarus, 2, 440

\bibitem[{{Harrington}(1968)}]{harrington68}
{Harrington}, R.~S. 1968, \aj, 73, 190

\bibitem[{{Herwig}(2000)}]{herwig00}
{Herwig}, F. 2000, \aap, 360, 952

\bibitem[{{Hills} \& {Day}(1976)}]{hills76}
{Hills}, J.~G., \& {Day}, C.~A. 1976, \aplett, 17, 87

\bibitem[{{H{\"o}fner}(2009)}]{hofner09}
{H{\"o}fner}, S. 2009, in Astronomical Society of the Pacific Conference
  Series, Vol. 414, Cosmic Dust - Near and Far, ed. T.~{Henning},
  E.~{Gr{\"u}n}, \& J.~{Steinacker}, 3

\bibitem[{{Hurley} {et~al.}(2002){Hurley}, {Tout}, \& {Pols}}]{hurley02}
{Hurley}, J.~R., {Tout}, C.~A., \& {Pols}, O.~R. 2002, \mnras, 329, 897

\bibitem[{{Ivanova} {et~al.}(2013){Ivanova}, {Justham}, {Chen}, {De Marco},
  {Fryer}, {Gaburov}, {Ge}, {Glebbeek}, {Han}, {Li}, {Lu}, {Marsh},
  {Podsiadlowski}, {Potter}, {Soker}, {Taam}, {Tauris}, {van den Heuvel}, \&
  {Webbink}}]{ivanova13}
{Ivanova}, N. {et~al.} 2013, \aapr, 21, 59

\bibitem[{{Jiang} \& {Tremaine}(2010)}]{jiang10}
{Jiang}, Y.-F., \& {Tremaine}, S. 2010, \mnras, 401, 977

\bibitem[{{Johnson} \& {Sandage}(1955)}]{johnson55}
{Johnson}, H.~L., \& {Sandage}, A.~R. 1955, \apj, 121, 616

\bibitem[{{Kalirai} {et~al.}(2014){Kalirai}, {Marigo}, \&
  {Tremblay}}]{kalirai14}
{Kalirai}, J.~S., {Marigo}, P., \& {Tremblay}, P.-E. 2014, \apj, 782, 17

\bibitem[{{Karovicova} {et~al.}(2013){Karovicova}, {Wittkowski}, {Ohnaka},
  {Boboltz}, {Fossat}, \& {Scholz}}]{karovicova13}
{Karovicova}, I., {Wittkowski}, M., {Ohnaka}, K., {Boboltz}, D.~A., {Fossat},
  E., \& {Scholz}, M. 2013, \aap, 560, A75

\bibitem[{{Kiseleva} {et~al.}(1998){Kiseleva}, {Eggleton}, \&
  {Mikkola}}]{kiseleva98}
{Kiseleva}, L.~G., {Eggleton}, P.~P., \& {Mikkola}, S. 1998, \mnras, 300, 292

\bibitem[{{Koester} {et~al.}(2014){Koester}, {G{\"a}nsicke}, \&
  {Farihi}}]{koester14}
{Koester}, D., {G{\"a}nsicke}, B.~T., \& {Farihi}, J. 2014, \aap, 566, A34

\bibitem[{{Koester} {et~al.}(2009){Koester}, {Voss}, {Napiwotzki},
  {Christlieb}, {Homeier}, {Lisker}, {Reimers}, \& {Heber}}]{koester09}
{Koester}, D., {Voss}, B., {Napiwotzki}, R., {Christlieb}, N., {Homeier}, D.,
  {Lisker}, T., {Reimers}, D., \& {Heber}, U. 2009, \aap, 505, 441

\bibitem[{{Kozai}(1962)}]{kozai62}
{Kozai}, Y. 1962, \aj, 67, 591

\bibitem[{{Kurtz} \& {Mink}(1998)}]{kurtz98}
{Kurtz}, M.~J., \& {Mink}, D.~J. 1998, \pasp, 110, 934

\bibitem[{{Kushnir} {et~al.}(2013){Kushnir}, {Katz}, {Dong}, {Livne}, \&
  {Fern{\'a}ndez}}]{kushnir13}
{Kushnir}, D., {Katz}, B., {Dong}, S., {Livne}, E., \& {Fern{\'a}ndez}, R.
  2013, \apjl, 778, L37

\bibitem[{{Lidov}(1962)}]{lidov62}
{Lidov}, M.~L. 1962, \planss, 9, 719

\bibitem[{{Lombardi} {et~al.}(2002){Lombardi}, {Warren}, {Rasio}, {Sills}, \&
  {Warren}}]{lombardi02}
{Lombardi}, Jr., J.~C., {Warren}, J.~S., {Rasio}, F.~A., {Sills}, A., \&
  {Warren}, A.~R. 2002, \apj, 568, 939

\bibitem[{{Marigo} {et~al.}(2013){Marigo}, {Bressan}, {Nanni}, {Girardi}, \&
  {Pumo}}]{marigo13}
{Marigo}, P., {Bressan}, A., {Nanni}, A., {Girardi}, L., \& {Pumo}, M.~L. 2013,
  \mnras, 434, 488

\bibitem[{{Marsh} {et~al.}(2011){Marsh}, {G{\"a}nsicke}, {Steeghs},
  {Southworth}, {Koester}, {Harris}, \& {Merry}}]{marsh11}
{Marsh}, T.~R., {G{\"a}nsicke}, B.~T., {Steeghs}, D., {Southworth}, J.,
  {Koester}, D., {Harris}, V., \& {Merry}, L. 2011, \apj, 736, 95

\bibitem[{{Mastrodemos} \& {Morris}(1999)}]{mastrodemos99}
{Mastrodemos}, N., \& {Morris}, M. 1999, \apj, 523, 357

\bibitem[{{Maxted} {et~al.}(2000){Maxted}, {Marsh}, {Moran}, \&
  {Han}}]{maxted00}
{Maxted}, P.~F.~L., {Marsh}, T.~R., {Moran}, C.~K.~J., \& {Han}, Z. 2000,
  \mnras, 314, 334

\bibitem[{{Mazeh} \& {Shaham}(1979)}]{mazeh79}
{Mazeh}, T., \& {Shaham}, J. 1979, \aap, 77, 145

\bibitem[{{McCrea}(1964)}]{mccrea64}
{McCrea}, W.~H. 1964, \mnras, 128, 147

\bibitem[{{Mohamed} \& {Podsiadlowski}(2007)}]{mohamed07}
{Mohamed}, S., \& {Podsiadlowski}, P. 2007, in Astronomical Society of the
  Pacific Conference Series, Vol. 372, 15th European Workshop on White Dwarfs,
  ed. R.~{Napiwotzki} \& M.~R. {Burleigh}, 397

\bibitem[{{Mohamed} \& {Podsiadlowski}(2012)}]{mohamed12}
{Mohamed}, S., \& {Podsiadlowski}, P. 2012, Baltic Astronomy, 21, 88

\bibitem[{{Naoz}(2016)}]{naoz16}
{Naoz}, S. 2016, ArXiv e-prints

\bibitem[{{Naoz} \& {Fabrycky}(2014)}]{naoz14}
{Naoz}, S., \& {Fabrycky}, D.~C. 2014, \apj, 793, 137

\bibitem[{{Naoz} {et~al.}(2011){Naoz}, {Farr}, {Lithwick}, {Rasio}, \&
  {Teyssandier}}]{naoz11}
{Naoz}, S., {Farr}, W.~M., {Lithwick}, Y., {Rasio}, F.~A., \& {Teyssandier}, J.
  2011, \nat, 473, 187

\bibitem[{{Naoz} {et~al.}(2012){Naoz}, {Farr}, \& {Rasio}}]{naoz12}
{Naoz}, S., {Farr}, W.~M., \& {Rasio}, F.~A. 2012, \apjl, 754, L36

\bibitem[{{Naoz} {et~al.}(2013){Naoz}, {Kocsis}, {Loeb}, \& {Yunes}}]{naoz13b}
{Naoz}, S., {Kocsis}, B., {Loeb}, A., \& {Yunes}, N. 2013, \apj, 773, 187

\bibitem[{{Nebot G{\'o}mez-Mor{\'a}n} {et~al.}(2011){Nebot
  G{\'o}mez-Mor{\'a}n}, {G{\"a}nsicke}, {Schreiber}, {Rebassa-Mansergas},
  {Schwope}, {Southworth}, {Aungwerojwit}, {Bothe}, {Davis}, {Kolb},
  {M{\"u}ller}, {Papadaki}, {Pyrzas}, {Rabitz}, {Rodr{\'{\i}}guez-Gil},
  {Schmidtobreick}, {Schwarz}, {Tappert}, {Toloza}, {Vogel}, \&
  {Zorotovic}}]{nebot11}
{Nebot G{\'o}mez-Mor{\'a}n}, A. {et~al.} 2011, \aap, 536, A43

\bibitem[{{Nomoto} {et~al.}(2007){Nomoto}, {Saio}, {Kato}, \&
  {Hachisu}}]{nomoto07}
{Nomoto}, K., {Saio}, H., {Kato}, M., \& {Hachisu}, I. 2007, \apj, 663, 1269

\bibitem[{{O'Brien} {et~al.}(2001){O'Brien}, {Bond}, \& {Sion}}]{obrien01}
{O'Brien}, M.~S., {Bond}, H.~E., \& {Sion}, E.~M. 2001, \apj, 563, 971

\bibitem[{{Olofsson} {et~al.}(2002){Olofsson}, {Gonz{\'a}lez Delgado},
  {Kerschbaum}, \& {Sch{\"o}ier}}]{olofsson02}
{Olofsson}, H., {Gonz{\'a}lez Delgado}, D., {Kerschbaum}, F., \& {Sch{\"o}ier},
  F.~L. 2002, \aap, 391, 1053

\bibitem[{{Paxton} {et~al.}(2011){Paxton}, {Bildsten}, {Dotter}, {Herwig},
  {Lesaffre}, \& {Timmes}}]{paxton11}
{Paxton}, B., {Bildsten}, L., {Dotter}, A., {Herwig}, F., {Lesaffre}, P., \&
  {Timmes}, F. 2011, \apjs, 192, 3

\bibitem[{{Paxton} {et~al.}(2013){Paxton}, {Cantiello}, {Arras}, {Bildsten},
  {Brown}, {Dotter}, {Mankovich}, {Montgomery}, {Stello}, {Timmes}, \&
  {Townsend}}]{paxton13}
{Paxton}, B. {et~al.} 2013, \apjs, 208, 4

\bibitem[{{Perets} \& {Fabrycky}(2009)}]{perets09}
{Perets}, H.~B., \& {Fabrycky}, D.~C. 2009, \apj, 697, 1048

\bibitem[{{Pietrinferni} {et~al.}(2004){Pietrinferni}, {Cassisi}, {Salaris}, \&
  {Castelli}}]{pietrinferni04}
{Pietrinferni}, A., {Cassisi}, S., {Salaris}, M., \& {Castelli}, F. 2004, \apj,
  612, 168

\bibitem[{{Pietrinferni} {et~al.}(2006){Pietrinferni}, {Cassisi}, {Salaris}, \&
  {Castelli}}]{pietrinferni06}
---. 2006, \apj, 642, 797

\bibitem[{{Raghavan} {et~al.}(2010){Raghavan}, {McAlister}, {Henry}, {Latham},
  {Marcy}, {Mason}, {Gies}, {White}, \& {ten Brummelaar}}]{raghavan10}
{Raghavan}, D. {et~al.} 2010, \apjs, 190, 1

\bibitem[{{Reipurth} \& {Mikkola}(2012)}]{reipurth12}
{Reipurth}, B., \& {Mikkola}, S. 2012, \nat, 492, 221

\bibitem[{{Sandage}(1953)}]{sandage53}
{Sandage}, A.~R. 1953, \aj, 58, 61

\bibitem[{{Sanduleak} \& {Pesch}(1982)}]{sanduleak82}
{Sanduleak}, N., \& {Pesch}, P. 1982, \iaucirc, 3703, 1

\bibitem[{{Santander-Garc{\'{\i}}a} {et~al.}(2015){Santander-Garc{\'{\i}}a},
  {Rodr{\'{\i}}guez-Gil}, {Corradi}, {Jones}, {Miszalski}, {Boffin},
  {Rubio-D{\'{\i}}ez}, \& {Kotze}}]{santander15}
{Santander-Garc{\'{\i}}a}, M., {Rodr{\'{\i}}guez-Gil}, P., {Corradi}, R.~L.~M.,
  {Jones}, D., {Miszalski}, B., {Boffin}, H.~M.~J., {Rubio-D{\'{\i}}ez}, M.~M.,
  \& {Kotze}, M.~M. 2015, \nat, 519, 63

\bibitem[{{Sills} {et~al.}(2005){Sills}, {Adams}, \& {Davies}}]{sills05}
{Sills}, A., {Adams}, T., \& {Davies}, M.~B. 2005, \mnras, 358, 716

\bibitem[{{Sion} {et~al.}(1991){Sion}, {Oswalt}, {Liebert}, \&
  {Hintzen}}]{sion91}
{Sion}, E.~M., {Oswalt}, T.~D., {Liebert}, J., \& {Hintzen}, P. 1991, \aj, 101,
  1476

\bibitem[{{Skrutskie} {et~al.}(2006){Skrutskie}, {Cutri}, {Stiening},
  {Weinberg}, {Schneider}, {Carpenter}, {Beichman}, {Capps}, {Chester},
  {Elias}, {Huchra}, {Liebert}, {Lonsdale}, {Monet}, {Price}, {Seitzer},
  {Jarrett}, {Kirkpatrick}, {Gizis}, {Howard}, {Evans}, {Fowler}, {Fullmer},
  {Hurt}, {Light}, {Kopan}, {Marsh}, {McCallon}, {Tam}, {Van Dyk}, \&
  {Wheelock}}]{2mass}
{Skrutskie}, M.~F. {et~al.} 2006, \aj, 131, 1163

\bibitem[{{Theuns} {et~al.}(1996){Theuns}, {Boffin}, \& {Jorissen}}]{theuns96}
{Theuns}, T., {Boffin}, H.~M.~J., \& {Jorissen}, A. 1996, \mnras, 280, 1264

\bibitem[{{Tokovinin} {et~al.}(2006){Tokovinin}, {Thomas}, {Sterzik}, \&
  {Udry}}]{tokovinin06}
{Tokovinin}, A., {Thomas}, S., {Sterzik}, M., \& {Udry}, S. 2006, \aap, 450,
  681

\bibitem[{{Tokovinin} \& {Smekhov}(2002)}]{tokovinin02}
{Tokovinin}, A.~A., \& {Smekhov}, M.~G. 2002, \aap, 382, 118

\bibitem[{{Tremblay} \& {Bergeron}(2009)}]{tremblay09}
{Tremblay}, P.-E., \& {Bergeron}, P. 2009, \apj, 696, 1755

\bibitem[{{Tremblay} {et~al.}(2011){Tremblay}, {Bergeron}, \&
  {Gianninas}}]{tremblay11}
{Tremblay}, P.-E., {Bergeron}, P., \& {Gianninas}, A. 2011, \apj, 730, 128

\bibitem[{{Tremblay} {et~al.}(2010){Tremblay}, {Bergeron}, {Kalirai}, \&
  {Gianninas}}]{tremblay10}
{Tremblay}, P.-E., {Bergeron}, P., {Kalirai}, J.~S., \& {Gianninas}, A. 2010,
  \apj, 712, 1345

\bibitem[{{Vanderplas} {et~al.}(2012){Vanderplas}, {Connolly}, {Ivezi{\'c}}, \&
  {Gray}}]{astroML}
{Vanderplas}, J., {Connolly}, A., {Ivezi{\'c}}, {\v Z}., \& {Gray}, A. 2012, in
  Conference on Intelligent Data Understanding (CIDU), 47 --54

\bibitem[{{Vassiliadis} \& {Wood}(1993)}]{vassiliadis93}
{Vassiliadis}, E., \& {Wood}, P.~R. 1993, \apj, 413, 641

\bibitem[{{Weiss} \& {Ferguson}(2009)}]{weiss09}
{Weiss}, A., \& {Ferguson}, J.~W. 2009, \aap, 508, 1343

\bibitem[{{Williams} {et~al.}(2009){Williams}, {Bolte}, \&
  {Koester}}]{williams09}
{Williams}, K.~A., {Bolte}, M., \& {Koester}, D. 2009, \apj, 693, 355

\bibitem[{{Wolf} {et~al.}(2013){Wolf}, {Bildsten}, {Brooks}, \&
  {Paxton}}]{wolf13}
{Wolf}, W.~M., {Bildsten}, L., {Brooks}, J., \& {Paxton}, B. 2013, \apj, 777,
  136

\bibitem[{{Wood}(1995)}]{wood95}
{Wood}, M.~A. 1995, in Lecture Notes in Physics, Berlin Springer Verlag, Vol.
  443, White Dwarfs, ed. D.~{Koester} \& K.~{Werner}, 41

\bibitem[{{Wright} {et~al.}(2010){Wright}, {Eisenhardt}, {Mainzer}, {Ressler},
  {Cutri}, {Jarrett}, {Kirkpatrick}, {Padgett}, {McMillan}, {Skrutskie},
  {Stanford}, {Cohen}, {Walker}, {Mather}, {Leisawitz}, {Gautier}, {McLean},
  {Benford}, {Lonsdale}, {Blain}, {Mendez}, {Irace}, {Duval}, {Liu}, {Royer},
  {Heinrichsen}, {Howard}, {Shannon}, {Kendall}, {Walsh}, {Larsen}, {Cardon},
  {Schick}, {Schwalm}, {Abid}, {Fabinsky}, {Naes}, \& {Tsai}}]{wise}
{Wright}, E.~L. {et~al.} 2010, \aj, 140, 1868

\bibitem[{{York} {et~al.}(2000){York}, {Adelman}, {Anderson}, {Anderson},
  {Annis}, {Bahcall}, {Bakken}, {Barkhouser}, {Bastian}, {Berman}, {Boroski},
  {Bracker}, {Briegel}, {Briggs}, {Brinkmann}, {Brunner}, {Burles}, {Carey},
  {Carr}, {Castander}, {Chen}, {Colestock}, {Connolly}, {Crocker}, {Csabai},
  {Czarapata}, {Davis}, {Doi}, {Dombeck}, {Eisenstein}, {Ellman}, {Elms},
  {Evans}, {Fan}, {Federwitz}, {Fiscelli}, {Friedman}, {Frieman}, {Fukugita},
  {Gillespie}, {Gunn}, {Gurbani}, {de Haas}, {Haldeman}, {Harris}, {Hayes},
  {Heckman}, {Hennessy}, {Hindsley}, {Holm}, {Holmgren}, {Huang}, {Hull},
  {Husby}, {Ichikawa}, {Ichikawa}, {Ivezi{\'c}}, {Kent}, {Kim}, {Kinney},
  {Klaene}, {Kleinman}, {Kleinman}, {Knapp}, {Korienek}, {Kron}, {Kunszt},
  {Lamb}, {Lee}, {Leger}, {Limmongkol}, {Lindenmeyer}, {Long}, {Loomis},
  {Loveday}, {Lucinio}, {Lupton}, {MacKinnon}, {Mannery}, {Mantsch}, {Margon},
  {McGehee}, {McKay}, {Meiksin}, {Merelli}, {Monet}, {Munn}, {Narayanan},
  {Nash}, {Neilsen}, {Neswold}, {Newberg}, {Nichol}, {Nicinski}, {Nonino},
  {Okada}, {Okamura}, {Ostriker}, {Owen}, {Pauls}, {Peoples}, {Peterson},
  {Petravick}, {Pier}, {Pope}, {Pordes}, {Prosapio}, {Rechenmacher}, {Quinn},
  {Richards}, {Richmond}, {Rivetta}, {Rockosi}, {Ruthmansdorfer}, {Sandford},
  {Schlegel}, {Schneider}, {Sekiguchi}, {Sergey}, {Shimasaku}, {Siegmund},
  {Smee}, {Smith}, {Snedden}, {Stone}, {Stoughton}, {Strauss}, {Stubbs},
  {SubbaRao}, {Szalay}, {Szapudi}, {Szokoly}, {Thakar}, {Tremonti}, {Tucker},
  {Uomoto}, {Vanden Berk}, {Vogeley}, {Waddell}, {Wang}, {Watanabe},
  {Weinberg}, {Yanny}, {Yasuda}, \& {SDSS Collaboration}}]{york00}
{York}, D.~G. {et~al.} 2000, \aj, 120, 1579

\end{thebibliography}
\setlength{\baselineskip}{1.667\baselineskip}

\end{document}